\documentstyle[prl,aps,preprint,tighten]{revtex}
\begin{document}
\draft
\title{Charge Transport in Junctions between D-Wave Superconductors}

\author{Yu.S.Barash$^a$, A.V.Galaktionov$^a$, and A.D. Zaikin$^{a,b}$ \\}

\address{
 $a)$ I.E. Tamm Department of Theoretical Physics, P.N. Lebedev Physics
Institute \\
     Leninsky Prospect 53, Moscow 117924, Russia \\
 $b)$ Institut f\"ur Theoretische
 Festk\"orperphysik, Universit\"at Karlsruhe, \\
 76128 Karlsruhe,  FRG\\
}
\date{\today}
\maketitle
\begin{abstract}
We develop a microscopic analysis of superconducting and dissipative currents
in junctions between superconductors with d-wave symmetry of the order
parameter.
We study the proximity effect in such superconductors and show that for certain
crystal orientations the superconducting order parameter can be essentially
suppressed in the vicinity of a nontransparent specularly reflecting boundary.
This effect strongly influences the value and the angular dependence of the
dc Josephson current $j_S$. At $T \sim T_c $ it leads
to a crossover between $j_S \propto T_c-T$ and $j_S \propto (T_c-T)^2$
respectively
for homogeneous and nonhomogeneous distribution of the order parameter in
the vicinity of a tunnel junction. We show that at low temperatures the
current-phase relation $j_S(\varphi )$ for SNS junctions and short weak links
between d-wave superconductors is essentially nonharmonic and contains a
discontinuity at $\varphi =0$. This leads to
further interesting features of such systems which can be used for
novel pairing symmetry tests in high temperature superconductors (HTSC).
We also investigated the low temperature I-V curves of NS and SS tunnel
junctions
and demonstrated that depending on the junction type and crystal orientation
these curves show zero-bias anomalies $I\propto V^2$, $I\propto V^2\ln (1/V)$
and $I\propto V^3$ caused by the gapless behavior of the order parameter in
d-wave
superconductors. Many of our results agree well with recent experimental
findings
for HTSC compounds.

 \end{abstract}
 \narrowtext
\section{Introduction}

In spite of enormous efforts made to understand the physical mechanisms of
pairing in various high temperature superconductors (HTSC) the situation
still remains unclear. A key role in understanding of this phenomenon belongs
to the question about the symmetry of the order parameter. Quite early after
the original Bednords and M\"uller discovery \cite{BM} the symmetry
of the $d_{x^2-y^2}$-type was suggested for HTSC materials \cite{R,B,K,G}.
Since
then a plenty of experiments have been designed
to probe the symmetry of the order parameter in HTSC (see e.g. \cite{Dy} for a
review).  Although many experimental results are consistent with the picture of
d-wave pairing (e.g. the temperature dependence of the penetration depth
\cite{Hardy},
NMR and NQR studies \cite{Kit} etc.) they still do not allow to rule out other
possibilities, like anisotropic s-wave pairing. Moreover the results of some
other
experiments (see e.g. \cite{9}) may indicate s-wave rather than
d-wave symmetry of the order parameter in HTSC. Therefore it is quite likely
that only a set of different and independent experimental tests
would allow to make an unambiquous conclusion about the order parameter
symmetry
in HTSC compounds.

An important information about the symmetry of superconducting pairing can be
obtained from the measurements of both the dc Josephson effect and the
quasiparticle current in
tunnel junctions between two HTSC. The dc Josephson effect in unconventional
superconductors has been discussed by Geshkenbein, Larkin and Barone \cite{lar}
and by Sigrist and Rice \cite{sig} who demonstated that the d-wave symmetry of
the
order parameter may lead to the sign inversion of the Josephson critical
current
for certain crystal orientations. Under these conditions the tunnel junction
becomes
the so-called $\pi$-junction \cite{Bul}. Being closed by a $\pi$-junction a
SQUID loop
with a not very small inductance develops a spontaneous circulating current
\cite{Bul}.
As a result the magnetic flux equal to a half of the flux quantum occurs inside
a ring and can be easily measured. Measurements of that kind have been carried
out for HTSC
samples \cite{W,Ott,Kirtley} and indeed demonstrated the results fully
consistent with the above
picture. These results in combination with the paramagnetic behavior of
granular
HTSC compounds \cite{Braunisch} and its theoretical interpretation
\cite{sig,Kh,PZ}
serve as a serious argument in favour of d-wave pairing symmetry in HTSC.

In contrast to the Josephson effect which is sensitive to the order parameter
phase difference across the junction low temperature measurements of the
quasiparticle current
in tunnel junctions provide information about the quasiparticle
density of states in superconducting banks and allow to distinguish
gapless superconductivity from that with a finite gap. The results of numerous
experiments vary from a nearly BCS-like to a clear
gapless behavior for different HTSC materials (see e.g. \cite{Dy} and
references
therein). The results of recent tunneling experiments with
Bi$_2$Sr$_2$CaCu$_2$O$_8$ samples
\cite{Mand,Ved} indicate a gapless behavior of the I-V curve at low voltages
and temperatures
(e.g. the dependence $I \propto V^3$ at low $V$ was reported in \cite{Mand}).

A growing number of experimental data makes it necessary to develop a detailed
analysis and specify theoretical predictions concerning both dc Josephson
effect
and quasiparticle tunneling between d-wave superconductors. Several
phenomenological
calculations assuming d-wave pairing have been already done (see below)
attempting
partial understanding of some experimental results. Nevertheless a number of
important questions still has to be adressed in this context. E.g. a uniform
distribution of the superconducting order parameter on both sides of a tunnel
barrier was assumed in many calculations. Being obviously correct for isotropic
s-wave superconductors this assumption may fail for d-wave ones depending on
their orientation relative to the tunnel barrier plane. Below we will show that
the spatial dependence of the order parameter due to the proximity effect
becomes
particularly important close to the superconducting critical temperature $T_c$
having a strong impact on the dc Josephson effect in d-wave superconductors.

Another problem appears if one applies the tunneling Hamiltonian method
to investigation of the charge transport through tunnel junctions in the d-wave
case. The momentum dependence
of the matrix elements describing tunneling between superconductors becomes
particularly
important in this case. It is easy to show that the choice
of the tunneling matrix elements as being independent on the  momentum
direction
(standard for the s-wave case) leads to confusing results for d-wave
superconductors.
Furthermore, an unambiguous choice of this dependence cannot be done within
this method.
This emphasizes the necessity to provide a microscopic description of tunneling
between such superconductors based on matching of the electron propagators at
the tunnel barrier. This approach leaves no space for ambiquity and,
on top of that, it is not confined to the
case of low transparency  barriers but allows to study other types of
weak links with highly transparent interfaces.

In this paper we will
provide an extensive microscopic study of the charge transport in various types
of junctions between d-wave superconductors with BCS-like behavior
of the density of states. The paper is organized as follows. In Section 2 we
develop a detailed study of the proximity effect for d-wave
superconductor-insulator
and d-wave superconductor-normal metal structures. We investigate the spatial
dependence of the superconducting gap function for various crystal orientations
and temperatures and show that for particular orientations the gap at the
superconductor-insulator interface can be completely suppressed. The dc
Josephson
current through tunnel junctions, SNS junctions and short weak links between
d-wave
superconductors is examined in Section 3. Our analysis allows to discover
several new qualitative features of the Josephson effect in such systems which
can
be used for further experimental tests of the pairing symmetry in HTSC. In
Section
4 we investigate the I-V curves for superconductor-superconductor (SS) and
normal
metal-superconductor (NS) tunnel junctions. For most of crystal orientations we
found gapless nonOhmic behavior in the limit of small voltages and $T=0$.
Discussion
of our results is presented in Section 5.

\section{Proximity Effect in d-Wave Superconductors}

The order parameter in bulk superconductors with unconventional pairing
depends on the direction of the Fermi momentum $\bbox{p_F}$
 \cite{volgor,su}.  Beyond that close to the edges of a
superconducting piece of metal the order parameter acquires a spatial
dependence due to the proximity effect. In this section we present a detailed
investigation of this effect for superconductors with d-wave symmetry of the
order
parameter. We show that the spatial dependence of the order parameter
in the vicinity
of a low transparency insulating barrier may essentially depend on the
crystal orientation  relative to the barrier plane. Similar --
although quantitatively different -- results hold provided a superconductor is
in a good electric contact with a normal metal.

In order to describe the proximity effect in d-wave superconductors we make use
of the Eilenberger equations for the quasiclassical Green functions \cite{eil}.
In the case of superconductors with singlet pairing these equations read
\cite{eil,rai1}
\begin{equation}
\left\{ \begin{array}{l}
(2\omega_{m}+ \bbox{ v}_{F} \nabla_{\bbox{ R}}) f(\hat{\bbox{ p}} , \bbox{ R},
\omega_{m}) - 2\Delta (\hat{\bbox{ p}}, \bbox{ R}) g(\hat{\bbox{ p}} , \bbox{
R}, \omega_{m})=0,\\ (2\omega_{m}- \bbox{ v}_{F} \nabla_{\bbox{ R}})
f^{+}(\hat{\bbox{ p}}, \bbox{ R}, \omega_{m}) - 2\Delta^{*} (\hat{\bbox{ p}},
\bbox{ R}) g(\hat{\bbox{ p}} , \bbox{ R}, \omega_{m})=0,\\ \bbox{ v}_{F}
 \nabla_{\bbox{ R}} g(\hat{\bbox{ p}} , \bbox{ R}, \omega_{m}) + \Delta
 (\hat{\bbox{ p}}, \bbox{ R}) f^{+}(\hat{\bbox{ p}}, \bbox{ R}, \omega_{m})
 -\Delta^{*} (\hat{\bbox{ p}}, \bbox{ R}) f(\hat{\bbox{ p}}, \bbox{ R},
 \omega_{m}) =0. \end{array} \right.
 \label{1} \end{equation} Here
 $\omega_{m}=(2m+1)\pi T$ is the Matsubara frequency, $\hat{\bbox{ p}}=
 \bbox{p_F}/|p_F|$,
 $\bbox{v}_{F}(\hat{\bbox{p}})=\bbox{p_F}/m$ is the Fermi velocity, $\Delta
 (\hat{\bbox{ p}}, \bbox{
 R})$ is the order parameter or the gap function. Anomalous and normal Green
functions
$ f(\hat{\bbox{ p}}, \bbox{ R}, \omega_{m})=f^{+*}(-\hat{\bbox{ p}} , \bbox{
R},
\omega_{m})$ and $ g(\hat{\bbox{ p}} , \bbox{ R}, \omega_{m})= g^{*}
 (-\hat{\bbox{ p}} , \bbox{ R}, \omega_{m}) $ obey the normalization condition
\begin{equation}
g^{2}(\hat{\bbox{ p}} , \bbox{
R}, \omega_{m})+ f(\hat{\bbox{ p}} , \bbox{ R}, \omega_{m}) f^{+}(\hat{\bbox{
p}} , \bbox{ R}, \omega_{m})= 1.  \label{2} \end{equation} The order parameter
$\Delta$ is linked to the anomalous Green function by means of the standard
selfconsistency equation \begin{equation} \Delta (\hat{\bbox{ p}}, \bbox{
R})=-\pi T\sum_{m} \int \frac{d^{2}S'}{(2\pi)^{3} v_{F}} V(\hat{\bbox{
p}},\hat{\bbox{ p'}})f(\hat{\bbox{ p'}} , \bbox{ R}, \omega_{m}) ,
\label{3}
\end{equation}
where $V(\hat{\bbox{ p}},\hat{\bbox{ p'}})$ is the anisotropic pairing
potential and the integration is carried out over the Fermi surface. The
quasiclassical equations (\ref{1}) are valid at the scale much larger than the
interatomic distance $\sim 1/p_F$
and do not keep track of rapid changes of the system parameters very close
to the
metal-metal or metal-insulator boundaries. In order to take these boundary
effects
into account the system of equations (\ref{1})-(\ref{3}) should
be supplemented by the boundary conditions \cite{zai,rai2,rai3} matching
the quasiclassical electron propagators $g$ and $f$ on both sides of the
boundary. These boundary conditions may essentially depend on the quality of
the
interface. In the case of a nonmagnetic specularly reflecting boundary between
two metals these conditions read \cite{zai}:
 $$ d_-(\hat{\bbox{
 p}}_-)=d_+(\hat{\bbox{ p}}_+),$$
 \begin{equation}
 d_-(\hat{\bbox{
 p}}_-)s_-^2(\hat{\bbox{ p}}_-)=\biggl[ (1+\frac{d_+(\hat{\bbox{ p}}_+)}{2})
 s_+(\hat{\bbox{ p}}_+),s_-(\hat{\bbox{ p}}_-)\biggr] \frac{1-R(\hat{\bbox{
 p}}_-)}{1+R(\hat{\bbox{ p}}_-)}.
  \label{4}
  \end{equation}
  Here $[a,b]$ denotes
the commutator of matrices $a$ and $b$, $R(\hat{\bbox{ p}})$ is the
 reflectivity coefficient and the index $+(-)$ labels the electron momentum in
 the right (left) halfspace with respect to the boundary plane. The 2x2
matrices $d$ and $s$ are defined by the equations $d(\hat{\bbox{p}})=\tilde
g(\hat{\bbox{p}})-\tilde g(\check{\bbox{p}}), s(\hat{\bbox{p}})=\tilde
g(\hat{\bbox{p}})+\tilde g(\check{\bbox{p}})$, where $\hat{\bbox{p}}
(\check{\bbox{p}})$ denotes the incident (reflected) electron momentum and
\begin{equation}
\tilde g(\hat{\bbox{p}}) = \left(  \begin{array}{cc} g(\hat{\bbox{p}}) &
if(\hat{\bbox{p}})\\ -if^+(\hat{\bbox{p}})& -g(\hat{\bbox{p}})
\end{array} \right).
\label{5}
\end{equation}
Provided the transparency of the tunnel barrier is equal to zero
$D \equiv 1-R=0$  the equations (\ref{4}) yield \cite{kul}
\begin{equation}
g_{\pm}(\hat{\bbox{p}})=g_{\pm}(\check{\bbox{p}}), \phantom{d}
f_{\pm}(\hat{\bbox{p}})=f_{\pm}(\check{\bbox{p}}), \phantom{d}
f_{\pm}^{+}(\hat{\bbox{p}})=f_{\pm}^{+}(\check{\bbox{p}}),
\label{6}
\end{equation}
where the Green functions are taken at the metal-insulator boundary.
In the opposite limiting case of a transparent boundary between two metals
$D=1$
the equations (\ref{4}) reduce to a simple continuity conditions for the
Green functions $g_+=g_-$ and $f_+=f_-$ at the boundary plane.

For the sake of definiteness let us assume that a d-wave superconductor
occupies
a half space $x>0$. Provided there is no current flow in this superconductor
one
can choose the gap function $\Delta$ to be real there and define
$f_{1}=(f+f^{+})/2, \phantom{d} f_{2}=(f-f^{+})/2$. Then with the aid
of (\ref{1})
we find
\begin{equation}
f_{1}(\hat{\bbox{p}},x, \omega_{m}) -
\frac{v_{x}^{2}}{4\omega_{m}^{2}} \partial_{x}^{2} f_{1}(\hat{\bbox{p}},x,
\omega_{m}) - \frac{\Delta (\hat{\bbox{p}}, x)}{|\omega_{m}|} \{ 1 +
 \frac{v_{x}^{2}}{4\omega_{m}^{2}}( \partial_{x}
f_{1}(\hat{\bbox{p}},x, \omega_{m}))^{2} -f_{1}^{2}(\hat{\bbox{p}},x,
\omega_{m})\} ^{1/2}=0,
\label{7}
\end{equation}

Let us first consider the case of an impenetrable boundary situated at the
plane
$x=0$. Then the spatial dependence of the gap function $\Delta (x)$ is defined
by
the combination of the equations (\ref{1}), (\ref{3}) with the boundary
conditions
(\ref{6}) at $x=0$ and $f_1(x \rightarrow \infty )=f_{\infty}$,
 $\Delta (x \rightarrow \infty )=\Delta_{\infty}$, where
$f_{\infty}$ and $\Delta_{\infty}$ are the equilibrium values for the anomalous
Green function and the order parameter in the bulk superconductor. Provided
the order parameter obeys the condition
\begin{equation}
\Delta (\hat{\bbox{p}}, \bbox{ R})
=\Delta (\check{\bbox{p}}, \bbox{ R})
\label{unif}
\end{equation}
the solution of the above equations does not depend on $x$ and thus the
functions
$f_1$ and $\Delta$ coincide with their equilibrium
values far from the boundary.
For superconductors with d-wave symmetry of
the order parameter $\Delta (\bbox{p_F})$ of the
$(p_{x_0}^{2}-p_{y_0}^{2})$-type
the condition (\ref{unif}) is satisfied if one of the principal crystal
axes $x_0$, $y_0$ or $z_0$ is perpendicular to the boundary plane.

For other crystal orientations the order parameter
$\Delta (\hat{\bbox{p}}, x)$ turns out to be spacially inhomogeneous.
To proceed further let us assume that one can express the value $\Delta$
in the form $\Delta(\hat{\bbox{p}},x)=\psi
(\hat{\bbox{p}}) \eta (x)$
\cite{FN1}.  At temperatures close to $T_c$ and distances larger than the
correlation length $\xi_0$ from the boundary the function $\eta (x)$ obeys the
Ginzburg-Landau equations which have a well known solution \begin{equation}
\eta_{GL}(x)=\eta_\infty\tanh [(x+\beta)/\sqrt{2}\xi(T)], \label{GLsol}
\end{equation} $\eta_\infty$ is the equilibrium value of $\eta (x)$ far from
the boundary and $\xi(T)$ is the temperature dependent superconducting
coherence length. In the case of uniaxial symmetry we have
$\xi(T)=(\xi_{\parallel}^2(T)\cos^2\alpha+\xi_{\perp}^2(T)
\sin^2\alpha)^{1/2}$, where $\xi_{\parallel}(T)$ and $\xi_{\perp}(T)$ are the
values of the coherence length respectively in the basal plane and the
transversal
direction, $\alpha$ is the angle between the vector normal to the boundary and
 the basal plane.  The value of $\beta$ is defined by the boundary condition
$$q\eta'(0)=\eta(0)$$ and the parameter $q$ has to be derived from the
microscopic theory (Eqs.  (\ref{3}), (\ref{6}) and (\ref{7})).

In the vicinity of the critical temperature $T \sim T_{c}$
one can linearize the equation (\ref{7}) neglecting higher powers of $f_{1}$.
Then combining (\ref{6}) and (\ref{7}) one gets
\begin{equation}
f_{1} (\hat{\bbox{p}},x, \omega_{m})=\frac{1}{|v_{x}|} \int_{0}^{\infty}
\{ \exp(-|\frac{2\omega_{m}}{v_{x}}(x-x')|)\Delta(\hat{\bbox{p}},x')+
\exp(-|\frac{2\omega_{m}}{v_{x}}|(x+x'))\Delta(\check{\bbox{p}},x'))\} dx'.
\label{8}
\end{equation}
Substituting (\ref{8}) into (\ref{3}) and
setting $\Delta(\hat{\bbox{p}},x)=\psi
(\hat{\bbox{p}}) \eta (x)$ we arrive at the integral equation
$$\eta(x)=\frac{\pi T \lambda}{\int \psi^{2}(\hat{\bbox{p}})
d^{2}S}  \sum_{m} \int_{0}^{\infty} \int
d^{2}S \eta (x') \psi (\hat{\bbox{p}})
\{ \frac{\psi(\hat{\bbox{p}})}{|v_{x}|}
\exp(-|\frac{2\omega_{m}}{v_{x}}(x-x')|)+$$
\begin{equation} +\frac{\psi(\check{\bbox{
p}})}{|v_{x}|} \exp(-|\frac{2\omega_{m}}{v_{x}}|(x+x')) \} dx'.
\label{9}
 \end{equation}
The effective coupling constant for an
anisotropic superconductor  $\lambda$ is defined by the equation
 \begin{equation}
\lambda \psi (\hat{\bbox{p}})=-\int V(\hat{\bbox{p}},\hat{\bbox{p'}})
\psi(\hat{\bbox{p'}}) \frac{d^{2}S'}{(2\pi)^{3} v_{F}},
\label{10}
\end{equation}
which yields $\pi T_c\lambda \sum_{m} |\omega_{m}|^{-1}=1$.
The equation (\ref{9}) describes the behavior of $\eta (x)$
 at distances  $x \lesssim \xi (T)$ from the boundary.
This equation coincides with that derived in \onlinecite{sha} within the
framework of a different technique for the case of a small gap anisotropy.

 A trivial combination of the above equations also allows to evaluate the Green
 functions at $T$ close to $T_c$. E.g. the functions $f_1$ and $f_2$ at the
 superconductor-insulator boundary $x=0$ and $T\rightarrow T_{c}$ read
 $$
f _{1} (\hat{\bbox{p}}, \omega_{m})=\frac{\psi(\hat{\bbox{p}})+
\psi(\check{\bbox{p}})}{|v_{x}|}\int_{0}^{\infty} \exp
(-|\frac{2\omega_{m}}{v_{x}}|x)\eta (x) dx, $$ \begin{equation} f _{2}
(\hat{\bbox{p}}, \omega_{m})=\frac{\psi(\check{\bbox{p}})-
\psi(\hat{\bbox{p}})}{v_{x}}\mbox{sgn}\, \omega_{m} \int_{0}^{\infty} \exp
(-|\frac{2\omega_{m}}{v_{x}}|x)\eta (x) dx
\label{11}
\end{equation}
$$=
\frac{\psi(\check{\bbox{p}})-\psi(\hat{\bbox{p}})}{\psi(\hat{\bbox{
p}})+\psi(\check{\bbox{p}})}\mbox{sgn}\, (v_{x} \omega_{m})
f _{1} (\hat{\bbox{p}}, \omega_{m}).
$$
The exact solution of (\ref{9}) can be easily found for the case
$\psi(\hat{\bbox{p}})=
-\psi(\check{\bbox{p}})$. Then we have
$$\eta (x)=Cx,$$ i.e. for this particular crystal orientation the gap function
vanishes at the boundary $x=0$ and for any $x>0$ it is described by the
function
$$\eta (x)=\eta_{\infty}\tanh [x/\sqrt{2}\xi(T)].$$ Combining the equation
$\psi(\hat{\bbox{p}})=-\psi(\check{\bbox{p}})$  with the symmetry condition for
the order parameter of the $(p_{x_0}^2-p_{y_0}^2)$-type we come to the
conclusion that the gap function vanishes at the superconductor-insulator
interface
provided  the principal crystal axes $x_0$ and $y_0$ constitutes the
angle $\pi /4$ with this interface.
Note that this conclusion is essentially based on the
symmetry arguments and thus remains valid at any temperature
below $T_{c}$. Indeed, for a pairing potential with the
symmetry property $V(\hat{\bbox{p}}, \hat{\bbox{p}}')=- V(\hat{\bbox{p}},
\check{\bbox{ p}}')$ and the boundary condition $f(\hat{\bbox{p}})=
f(\check{\bbox{p}})$ we obtain $\eta (0)=0$  from the selfconcistency
equation (\ref{3}) at any $T$.

At distances $x\gg \xi_{0}$ from the boundary the integral equation (\ref{9})
has a general solution
\begin{equation}
\eta (x)= C(x+q).
\label{sol}
\end{equation}
The parameter $q$ can be easily found from a simple variational procedure
\cite{svi} which yields
$$
q= \biggl( \frac{\pi^{3}}{336\zeta(3)T}
\int_{v_{x}>0}
[\psi(\hat{\bbox{p}})+\psi(\check{\bbox{p}})]^{2}v_{x}^{3}d^{2}S+
\frac{7\zeta(3)}{4\pi^{3}T}
\frac{(
\int_{v_{x}>0} [\psi(\hat{\bbox{p}})+\psi(\check{\bbox{p}})]^{2}v_{x}^{2}d^{2}S
)^{2}}{
\int_{v_{x}>0} [\psi(\hat{\bbox{p}})-\psi(\check{\bbox{p}})]^{2}v_{x}d^{2}S
}\biggr)$$\begin{equation}
( \int\psi^{2}(\hat{\bbox{p}})v_{x}^{2}d^{2}S)^{-1},
\label{12}
\end{equation}
where integration over the Fermi surface is confined to its part with
$v_{x}>0$.

Note that with the aid of general symmetry arguments one can unambiguously
fix only those crystal orientations for which the parameter $q$ is equal to
zero and infinity.
The detailed form of $q$ as a function of a crystal orientation relative to
the boundary plane for a given pairing symmetry type
 essentially depends on the particular choice of the basis
function $\psi(\hat{\bbox{p}})$. E.g. for the pairing symmetry of the type
$(p_{x_0}^2-p_{y_0}^2)$ for the sake of definiteness one can choose
the simplest basis function $\psi(\hat{\bbox{
p}})=\Delta_0(\hat p_{x_0}^2-\hat p_{y_0}^2)$ and after an explicit integration
in (\ref{12}) get
$$
q=\frac{4\zeta(3)v_F}{5\pi^3 T_c (1+2\sin^2\theta_0)}\biggl\{
\frac{(\cos^2\theta_0 + 3\sin^4\theta_0\cos^2 2\phi_0)^2}{\sin^2\theta_0
(1-\sin^2\theta_0\cos^2 2\phi_0)}+$$ \begin{equation}
+\frac{25}{6\zeta^2(3)}\biggl(\frac{\pi}{4}\biggr)^6(4\cos^2\theta_0 + 19
\sin^4\theta_0\cos^2 2\phi_0)\biggr\} .
\label{13} \end{equation}
Here
$\theta_0,\ \phi_0$ are the polar and the azimuthal angles of the normal
$\bbox{ n}$ to the boundary (see Fig.1). The function $q(\theta_0,\varphi_0)$
is
presented in Fig.2. We see that the parameter $q$ varies
from $q=0$ (or $\eta =0$) for $\theta_0=\pi/2$, $\phi_0=\pi/4$
($\psi(\hat{\bbox{p}})= -\psi(\check{\bbox{ p}})$) to $q=\infty$ (i.e.
$\eta'(0)=0$) if one of the main crystal axes $x_0$, $
y_0$ or $ z_0$ has the angle $\pi /2$ with the interface plane.

Our analysis can be easily modified to take into account the effect of
nonmagnetic impurities. In the presence of such impurities
close to $T_c$ the
equation (\ref{9}) remains valid if one substitutes
$\omega_{m} \rightarrow \tilde{\omega}_{m}
=\omega_{m}+\mbox{sgn}\,\omega_{m}/2\tau_{imp}$, where $\tau_{imp}$ is the
average scattering time.
Accordingly the modified critical temperature $T'_{c}$ is defined by the
equation $\pi T'_{c} \lambda \sum_{m} |\tilde{\omega}_{m}|^{-1}=1$. With the
aid
of (\ref{9}) it is easy to check that scattering
on nonmagnetic impurities does not lead to qualitative changes in the behavior
of the order parameter in the vicinity of a superconductor-insulator interface.
E.g. the homogeneous solution $\eta (x)=\eta_{\infty}$ for
$\psi(\hat{\bbox{p}})=\psi(\check{\bbox{p}})$
and the solution $\eta(x)=Cx$ for
$\psi(\hat{\bbox{p}})=-\psi(\check{\bbox{p}})$
remain valid in the presence of impurities. Similar results hold also for
intermediate crystal orientations.

The above analysis shows that in the case of specularly reflecting boundaries
the values  $\eta(0)$ and $\eta_\infty$ are of the same order of magnitude only
for particular orientations  of the normal $\bbox{ n}$  within narrow
angular intervals $\Delta\phi_0\sim \Delta\theta_0\sim (\xi_0/\xi(T))^{1/2}$
around the crystal axes $x_0, y_0, z_0$. For other crystal orientations from
the equation (\ref{13}) one has
$\eta(0)\sim\eta_\infty\xi_0/\xi(T)\ll \eta_\infty$, i.e. the d-wave
superconducting order parameter turns out to be strongly suppressed in the
vicinity of the insulating barrier.
In the case of diffusive scattering at the interface one has
$q\sim \xi_{0}$ \cite{sha}. Thus in this case the parameter $\eta (0)\sim
\xi_{0}\eta_{\infty}/\xi (T)\ll \eta_{\infty}$ for $T\rightarrow T_{c}$ and
all crystal orientations.

At low temperatures the analysis of the proximity effect becomes more
difficult.
As we already discussed for particular crystal orientations
the form of the order parameter can be described with the aid of symmetry
arguments.
E.g. for $\psi(\hat{\bbox{p}})=\psi(\check{\bbox{p}})$ the superconducting
order
parameter
$\Delta (T)$ does not depend on coordinates whereas for
$\psi(\hat{\bbox{p}})=-\psi(\check{\bbox{p}})$ at any $T$ we have $\Delta =0$
at
the superconductor-insulator interface.  For other crystal orientations the
behavior of the order parameter at $T \ll T_c$ can be qualitatively described
by the following estimate.

Let us consider the exact equation (\ref{7}) and split the frequency range into
two intervals: $|\omega_m|\ll \Delta_{0}$ and
$|\omega_m| \gtrsim \Delta_{0}$.
The contribution of the first frequency interval to the selfconsistency
equation is small in the parameter $\omega_m/\Delta_0 $. Therefore for our
estimate it is sufficient to restrict our consideration to the second
frequency interval. For $|\omega_m|\gg \Delta_0 $
one can neglect nonlinear terms in (\ref{7}). Then making use of the
condition $\xi_0\gg v_x/\omega_m$ one gets
\begin{equation}
f _{1} (\hat{\bbox{p}},x, \omega_{m})=\frac{\eta(x)\psi(\hat{\bbox{
p}})}{|\omega_m|}+ \frac{\eta(0)[\psi(\check{\bbox{p}})-
\psi(\hat{\bbox{p}})]}{2|\omega_m|}\exp(-|\frac{2\omega_m}{v_x}|x).
\label{14}
\end{equation}
Taking (\ref{14}) as an approximate form for $f_1$ in the whole frequency
interval
$\omega_m\gtrsim\Delta_0$,  setting $\eta(x)=$const and expanding in
powers of the anisotropy parameter
$\int_{v_x>0}
[\psi(\hat{\bbox{p}})- \psi(\check{\bbox{p}})]^2d^2S/ \int\psi^2(\bbox{
p})d^2S $  we find
 \begin{equation} \eta(0)\approx \biggl( 1-\frac{\int_{v_x>0}
[\psi(\hat{\bbox{p}})- \psi(\check{\bbox{p}})]^2d^2S}{2\int\psi^2(\bbox{
p})d^2S}\biggr) \eta_\infty .
\label{15}
\end{equation}
This estimate provides correct limits $\eta (0) =\eta _{\infty}$ for
$\psi(\hat{\bbox{p}})=\psi(\check{\bbox{p}})$ and $\eta (0)=0$
for $\psi(\hat{\bbox{p}})=- \psi(\check{\bbox{p}})$ and  qualitatively
describes the low temperature behavior of $\eta (0)$  for intermediate crystal
orientations. It demonstrates that at $T \ll T_c$ the
value $\eta (0)$ is of order $\eta_\infty$ for a relatively wide
angular interval $\Delta\phi_0, \Delta\theta_0 \sim 1$. The typical length
scale at which the value $\eta(x)$ changes from $\eta (0)$ at the boundary
to $\eta_\infty$ deep in the superconductor is of
order $\xi_0$.

  In the case of an ideally transmitting normal metal-superconductor
 interface $D(\hat{\bbox{
p}})=1$ quasiclassical propagators are continuous at
this interface: $g_-(x=0)=g_+(x=0)$ and $f_-(x=0)=f_+(x=0)$. Then imposing the
boundary conditions $f_1(x \rightarrow \infty )=f_{\infty}$, $\Delta (x
\rightarrow \infty )=\Delta_{\infty}$, $f_1 (-\infty)=0$ and assuming that the
order parameter is equal to zero in the normal metal $\Delta (x<0) \equiv 0$
one can repeat the above analysis and show that at $T$ close to $T_c$ the order
parameter is again described by the equations (\ref{GLsol}) and (\ref{sol}),
where \begin{equation} q= \biggl( \frac{\pi^{3}}{336\zeta(3)T} \int
\psi^2(\hat{\bbox{p}})|v_{x}|^{3}d^{2}S+ \frac{7\zeta(3)}{4\pi^{3}T} \frac{(
\int \psi^2(\hat{\bbox{p}})v_{x}^{2}d^{2}S )^{2}}{ \int \psi^2(\hat{\bbox{p}})
|v_{x}|d^{2}S }\biggr) ( \int\psi^{2}(\hat{\bbox{p}})v_{x}^{2}d^{2}S)^{-1}.
\label{16}
\end{equation}
According to this result for NS structures at $T \rightarrow T_c$
the parameter $q$ is of order $\xi_{0}$ for all crystal orientations.
As before at $T\rightarrow 0$  the order parameter changes from $\Delta (x=0)$
to
$\Delta (x=\infty )$ at distances of order $\xi_{0}$ from the NS boundary. To
estimate the value $\eta(0)$  for this temperature interval one can follow the
procedure developed in
\cite{zazh} for the case of isotropic s-wave superconductors. Then similarly to
\cite{zaik,zazh} one finds $\eta(0)\approx 0.5\eta_{\infty}$  at
$T\rightarrow 0$.  This estimate appears to hold for any crystal orientation.
It also agrees with the results of \onlinecite{bruder} in which the order
parameter of a d-wave superconductor has been calculated numerically for a
particular crystal orientation.

Finally let us note that at $T$ close to $T_c$ and distances $x \lesssim \xi_0$
from the interface we also expect an additional suppression of the order
parameter
with respect to that described by the equations (\ref{sol}) and (\ref{12}),
(\ref{16}). Indeed, the Ginzburg-Landau equation and its solution
(\ref{GLsol}) apply only at distances $x \gg \xi_0$ from the boundary.
For $x \lesssim \xi _0$ it is necessary to proceed within the framework of a
rigorous microscopic analysis \cite{zaik}. E.g.
one can show \cite{zaik} that the exact value $\eta(0)$ at
the boundary between a normal metal and an $s$-wave superconductors  is by
the factor  $\approx$ 1.4 smaller  and the exact
ratio $\eta'(0)/\eta(0)$ is by the factor $\approx$ 1.6 larger as compared to
the
corresponding values which follow from the standard Ginzburg-Landau analysis
(see e.g. \cite{RZaitsev}). We believe that similar
situation takes place also for superconductors with anisotropic pairing
considered here.

\section{Josephson Current for Anisotropic Superconductors}
\subsection{Tunnel Junctions}

Let us investigate the dc Josephson effect in tunnel junctions between two
d-wave
superconductors.  Assuming that the junction transparency is small
$D(\hat{\bbox{p}}) \ll 1$ one can proceed perturbatively and expand the
boundary
conditions (\ref{4}) in powers of $D(\hat{\bbox{p}})$. Keeping only the linear
terms
one gets
 \begin{equation}
 g_-(\hat{\bbox{p}}_{-})- g_-(\check{\bbox{p}}_{-})= \frac{1}{2}
 D(\hat{\bbox{p}}_-) (f_+(\hat{\bbox{p}}_{+})f_-^{+}(\hat{\bbox{p}}_{-})
- f_-(\hat{\bbox{p}}_{-})f_+^{+}(\hat{\bbox{p}}_{+})),
\label{18}
\end{equation}
where the functions  $f_\pm(\hat{\bbox{p}}_{\pm})$ and
$f_\pm^+(\hat{\bbox{p}}_{\pm})$ are the anomalous Green functions calculated on
both sides of the tunnel barrier for $D(\hat{\bbox{p}}) =0$. Substituting this
expression into the formula for the superconducting current
\begin{equation}
\bbox{ j}(\bbox{ R}) = -2\pi ieT \sum_{m} \int \frac{d^{2}S}{(2\pi)^{3} v_{F}}
 \bbox{ v}_{F} (\hat{\bbox{p}}) g(\hat{\bbox{p}} , \bbox{ R}, \omega_{m})
 \label{17} \end{equation} we arrive at the general expression for the
Josephson current \begin{equation} j_S=2\pi eT\sin\varphi\sum_{m}
\int_{v_{x}>0}
D(\hat{\bbox{p}}_-) v_{x}(\hat{\bbox{p}}_-)[f_{1-}(\hat{\bbox{
p}}_-)f_{1+}(\hat{\bbox{p}}_{+}) - f_{2-}(\hat{\bbox{p}}_{-})f_{2+}(\hat{\bbox{
p}}_{+})]\frac{d^2S_-}{2\pi^3 v_F},
\label{19}
\end{equation}
where $\varphi$ is the phase difference between two superconductors (i.e. the
gap is proportional to $\exp(i\varphi)\psi_+(\hat{\bbox{p}}_+)$ and
$\psi_-(\hat{\bbox{p}}_-)$ on the left and on the right respectively), $v_{x}$
is the Fermi velocity projection on the normal to the plane interface. The
functions $f_1$, $f_2$ are calculated for real $\psi_+(\hat{\bbox{p}}_+)$,
$\psi_-(\hat{\bbox{p}}_-)$.  Provided the functions
$f_\pm(\hat{\bbox{p}}_\pm
,\omega_{m})$ do not depend on space coordinates one can easily evaluate $j_S$
(\ref{19})
and get \begin{equation}
j_{S}=2\pi e T \sin\varphi \sum_{m} \int_{v_{x}>0}
\frac { \Delta_+ (\hat{\bbox{p}}_{+}) \Delta_- (\hat{\bbox{p}}_{-})
D(\hat{\bbox{p}}_{-}) v_{x}(\hat{\bbox{p}}_{-}) } {\sqrt{ \Delta_+^{2}
(\hat{\bbox{p}}_{+}) + \omega_{m}^{2}} \sqrt{ \Delta_-^{2} (\hat{\bbox{p}}_{-})
+ \omega_{m}^{2}} } \frac{d^{2}S_{-}}{(2\pi)^{3} v_{F}}.
\label{20}
\end{equation}
This result coincides with that obtained in \onlinecite{rai2}.
At low temperatures $T \ll \Delta$ it yields
\begin{equation}
j_{S}=4 e \sin\varphi  \int_{v_{x}>0}
\frac { \Delta_+ (\hat{\bbox{p}}_{+}) \Delta_- (\hat{\bbox{p}}_{-})}{
|\Delta_+ (\hat{\bbox{p}}_{+})|+ |\Delta_- (\hat{\bbox{p}}_{-})|}
K\biggl(\frac{|\Delta_+ (\hat{\bbox{p}}_{+})|-|\Delta_- (\hat{\bbox{p}}_{-})|}{
|\Delta_+ (\hat{\bbox{p}}_{+})|+|\Delta_- (\hat{\bbox{p}}_{-})|}\biggr)
D(\hat{\bbox{p}}_{-}) v_{x}(\hat{\bbox{p}}_{-})
\frac{d^{2}S_{-}}{(2\pi)^{3} v_{F}},
\label{AB}
\end{equation}
where $K(t)=\int_{0}^{\pi/2}(1-t^2\sin^2\phi)^{-1/2}d\phi $  is the complete
elliptic integral. At $\Delta_-(\hat{\bbox{p}}_{-}) \simeq
\Delta_+ (\hat{\bbox{p}}_{+}) \simeq \Delta (\hat{\bbox{p}})$ and any $T$ we
find from (\ref{20})
\begin{equation}
j_{S}=\pi e \sin\varphi  \int_{v_{x}>0}
\Delta (\hat{\bbox{p}}) \tanh \biggl{(} \frac{\Delta
(\hat{\bbox{p}})}{2T}\biggr{)}
D(\hat{\bbox{p}}) v_{x}(\hat{\bbox{p}})
\frac{d^{2}S}{(2\pi)^{3} v_{F}}.
\label{AB2}
\end{equation}
This result shows that the Josephson current between identical similarly
oriented d-wave superconductors at $T \ll \Delta$ is proportional to the
product $\Delta (\hat{\bbox{p}})D(\hat{\bbox{p}}) v_{x}(\hat{\bbox{p}})$
averaged over the momentum directions at the Fermi surface.

With the aid of a standard expression for the normal
state resistance of a tunnel junction
 \begin{equation} R_{N}^{-1}= 2e^2\int_{v_{x}>0} D(\hat{\bbox{p}}_{-})
 v_{x}(\hat{\bbox{p}}_{-}) \frac{d^{2}S_{-}}{(2\pi)^{3} v_{F}}
 \label{RN}
 \end{equation}
one can easily reduce the equations (\ref{AB}) and (\ref{AB2}) to the analogous
results for conventional superconductors provided the momentum dependence of
the
gap function $\Delta$ is neglected.

Close to the critical
temperature $T \sim T_{c}$ the expression (\ref{20}) reduces to
\begin{equation} j_{S}=\frac{\pi e \sin\varphi}{2T}  \int_{v_{x}>0} \Delta_+
(\hat{\bbox{p}}_{+}) \Delta_- (\hat{\bbox{p}}_{-}) D(\hat{\bbox{p}}_{-})
 v_{x}(\hat{\bbox{p}}_{-}) \frac{d^{2}S_{-}}{(2\pi)^{3} v_{F}} .
\label{21}
\end{equation}

As it was demonstrated in the previous section in the presence of a
nontransparent
insulating barrier the order parameter and the Green functions of a d-wave
superconductor do not depend on coordinates only provided one of the principal
axes is perpendicular to the barrier plane. For any other crystal
orientation the order parameter
$\Delta $ as well as $g$- and $f$-functions vary in space and the results
(\ref{20})-(\ref{AB2}), (\ref{21})
are no longer valid. To derive the corresponding generalization of
Eq. (\ref{21}) let us combine the exact formula (\ref{19}) with the results
obtained in Section 2. As for a spherical form of the Fermi surface the value
$D(\hat{\bbox{p}})$ depends only on $\hat p_x$, with the aid of
Eq. (\ref{11}) at $T$ close to $T_c$ we find
\begin{equation} j_S=8\pi eT\sin\varphi \sum_{m}
\int_{v_{x}>0} D(\hat{\bbox{p}}_{-}) v_{x}(\hat{\bbox{p}}_{-})
I_+(\hat{\bbox{p}}_+,\omega_m)
I_-(\hat{\bbox{ p}}_-,\omega_m) \frac{d^{2}S_{-}}{(2\pi)^{3} v_{F}}.
\label{22}
\end{equation}
 The functions $I_{\pm}$ read
 \begin{equation}
 I_{\pm}(\hat{\bbox{p}}_{\pm},\omega_m)=\frac{1}{|v_x|}\int_{0}^{\infty}
 \Delta_\pm (\hat{\bbox{p}}_{\pm},x) \exp(-|\frac{2\omega_m}{v_x}|x)dx,
 \label{23}
 \end{equation}
 $x$ is the distance from the interface.

The results (\ref{22}), (\ref{23}) show that the
expression for the Josepson current in d-wave superconductors is essentially
nonlocal in space:  the value $j_S$ is determined by the order parameter at
distances $\simeq \xi_0$ from the tunnel barrier. E.g. for a particular crystal
orientation $\psi(\hat{\bbox{p}})=-\psi(\check{\bbox{p}})$ at both sides of the
barrier close to $T_{c}$ we get \begin{equation} j_{S}= (\eta'(0))^2 \frac{\pi
 e \sin\varphi}{96T^3}  \int_{v_{x}>0} \psi_+(\hat{\bbox{p}}_{+}) \psi_-
  (\hat{\bbox{p}}_{-}) D(\hat{\bbox{p}}_{-}) v_{x}^3(\hat{\bbox{p}}_{-})
 \frac{d^{2}S_{-}}{(2\pi)^{3} v_{F}}.
\label{24}
\end{equation}
This result demonstrates that even if the order parameter of a d-wave
superconductor
vanishes at the interface $\eta(0)=0$ the Josephson current remains finite.
The nonlocality of $j_S$ (\ref{22}), (\ref{23}) results in different
temperature
dependence of the Josephson current for different crystal orientations. E.g.
for a
particular case $\psi(\hat{\bbox{p}})=\psi(\check{\bbox{p}})$ and $T$ close to
$T_c$
from (\ref{21}) we have
\begin{equation}
j_S\propto \eta^2(0)\propto T_c-T,
\label{21a}
\end{equation}
whereas for $\psi(\hat{\bbox{p}})=-\psi(\check{\bbox{p}})$ the result
 (\ref{24}) yields
 \begin{equation}
 j_S\propto \eta'^2(0)\propto (T_c-T)^2.
\label{24a}
\end{equation}

To evaluate the Josephson current for arbitrary crystal orientations let us
combine the results (\ref{GLsol}), (\ref{sol}) with the equations
(\ref{22}), (\ref{23}). Then
making use of (\ref{RN})
and assuming $D(\hat{\bbox{p}})\propto p_n^k/p_F^k$ ($k=0,2,...$) \cite{FN2} we
get
 $$
j_SR_N=\frac{\sin\varphi}{16eTp_F^2}\frac{\eta_\infty^2(k+2)}{\xi_+(T)\xi_-(T)}
\int_{v_{x}>0} \biggl(q_+q_-+\frac{7\zeta(3)}{2\pi^3T}
(q_+ +q_-)v_{x}(\hat{\bbox{p}}_-) +$$ \begin{equation}
+\frac{v_{x}^2(\hat{\bbox{
p}}_-)}{48T^2}\biggr) \psi_-(\hat{\bbox{p}}_-)\psi_+(\hat{\bbox{p}}_+)
\frac{v_{x}^{k+1}(\hat{\bbox{p}}_-)}{v_F^{k+1}} d^2S .
\label{29}
\end{equation}
Here $q_{\pm}$ are the values of the parameter $q$ (\ref{13}) on both sides of
the interface. The result (\ref{29}) is valid provided the parameters $q_{\pm}$
are not
very large $q_{\pm} \ll \xi_\pm (T) $. According to (\ref{13}) this implies
that
the direction of either one of the crystal principal axes should not be very
close to that normal to the interface   $\Delta \phi_0 ,\Delta \theta_0 \gg
(\xi_0 / \xi(T))^{1/2} $.  At  $\Delta \phi_0 ,\Delta \theta_0 \lesssim(\xi_0 /
\xi(T))^{1/2} $  the
expression for $j_S$ shows a crossover from (\ref{29}) to (\ref{21}).  Thus we
can conclude that at $T$ close to $T_c$ for a wide range of crystal
orientations the proximity effect strongly influences the dc Josepson
current in d-wave superconductors leading to the temperature dependence  $j_S
\propto (\eta_{\infty}/\xi (T))^2 \propto (T_{c}-T)^2$. Only if one of the
crystal principal axes is (nearly) perpendicular to the junction plane this
dependence changes and
becomes $j_S \propto T_c-T$.  At lower temperatures $T \lesssim \Delta (T)$ the
role of the proximity effect becomes less important and the expression
(\ref{20}) qualitatively describes the dc Josephson current for a wide range of
crystal orientations.

Sigrist and Rice
\cite{sig} suggested the following simple phenomenological expression for the
Josephson current between two tetragonal superconductors with
$p_{x_0}^{2}-p_{y_0}^{2}$ type of pairing near $T_{c}$:
\begin{equation} j_{S}=j_{0}\sin \varphi
(n^{2}_{+(x_0)}- n^{2}_{+(y_0)}) (n^{2}_{-(x_0)}- n^{2}_{-(y_0)}).
 \label{27}
\end{equation} Here $n_{\pm (i)}$ denotes the projection of the unit vector
normal to the boundary plane on the $i$ axis.

Since the basis function $\psi(\hat{\bbox{ p}})$ is not unique for a given
pairing symmetry type, the dependence of the Josephson current on the
orientation of the boundary plane relative to the crystal axes cannot be chosen
unambiguously
from the symmetry arguments only. Eq.(\ref{27}) presents the simplest
example for a particular angular dependence of the Josephson current consistent
with the pairing symmetry. For more complicated cases higher powers of $n_{\pm
(x,y)}$ can appear.

In the case of diffusive  scattering  at the boundary the
Josephson current  does not depend on the relative orientation of two
superconductors  and $\eta(0)\sim(\xi_0/\xi(T))^2\eta_\infty $ \cite{lar}.
If we put $\eta(0)\approx 0 $ , in accordance with
(\ref{24a}),(\ref{29}) at $T$ close to $T_c$ we have $ j_{0} \propto
(T_{c}-T)^{2}$.

As it follows from our analysis the dependence of $j_S$ on the relative
orientation of
superconductors becomes important for the specular scattering at the
insulating barrier.
 From Eq. (\ref{21}) one can easily
recover the dependence of the Josephson current $j_{S}$ on the angle
$\phi$ between the $z_0$-axes of two superconductors. E.g. if
$n_{\pm (x)}=1$ and $ D \propto p_x^2/p_F^2$ we have
\begin{equation}
j_{S}R_N=\frac{\Delta_{1} \Delta_{2}\sin\varphi}{96eT}
(9+\frac{1}{2} \cos 2\phi),
\label{J}
\end{equation}
where $\Delta_{1} ,\Delta_{2}$
are the maximum values of $\Delta(\hat{\bbox{p}}_{+}),\phantom{d}
\Delta(\hat{\bbox{p}}_{-})$. We see that in the case of Eq. (\ref{J})
the Josephson critical current is positive for all values of $\phi$
(0-contact).
However if, for instance, the $z_0$-axes
for both superconductors are perpendicular to the boundary plane the Eq.
(\ref{21}) yields
\begin{equation}
j_{S}\propto \cos 2\phi ,
\label{J2}
\end{equation}
 where $\phi$ is now the angle between the $x_0$-axes of the
superconductors. In the latter case the Josephson critical current turns out
to be negative ($\pi$-junction \cite{Bul}) for certain values of $\phi$.
Thus, rotating one of the crystals around its $z_0$-axis one can turn the
0-junction
 into the $\pi$-junction.  The latter
property remains also for lower temperatures $T\ll T_{c}$ in which case the
$\phi$-dependence of the Josephson current is more complicated.

Note that the dependence (\ref{J2}) permits to realize a very simple
configuration, for which three parts of {\it the same} superconductor form
a closed circuit with the odd number of $\pi$-junctions (see Fig.3). The
$z_0$-axes of the
grains coincide and are taken to be perpendicular to junction planes.
The angles between the $x_0$-axes of the first and second  as well as the
second
and the third grains are supposed to be equal to $\pi/6$. Then according to
(\ref{J2}) the corresponding junctions are of the 0-type. In contrast the
junction
between the third and the first grains turns out to be of the $\pi$-type
because
the angle $\phi$ is equal to $\pi/3$ for this junction.

\subsection{SNS Junctions}

Let us now consider the dc Josephson effect in planar structures
superconductor-normal metal-superconductor
(SNS). In the case of clean s-wave superconductors this effect was studied in
Refs. \cite{kul69,ishii}. Here we provide the generalization to the case of
d-wave superconductors with arbitrary orientations relative to the boundaries.

In order to evaluate the supercurrent in SNS junctions one has to solve the
Eilenberger equations (\ref{1}) in superconducting and normal regions and match
the solutions at NS interfaces with the aid of the boundary conditions
(\ref{4}).
Below we shall consider the case of transparent NS boundaries with
$D(\hat{\bbox{p}})=1$ and assume that the order parameter in the normal metal
is equal to zero $\Delta (-d/2 <x< d/2) \equiv 0$, $d$ is the thickness of a
normal layer between two d-wave superconductors which occupy two halfspaces
$x<-d/2$ and $x>d/2$.
To find the Green functions of superconducting banks we shall use the following
ansatz
\begin{equation}
\left\{ \begin{array}{l}
f_\pm(\hat{\bbox{p}}_\pm,x,\omega_m)=e^{i\varphi_\pm(\hat{\bbox{p}}_\pm)}
\mp e^{i\varphi_\pm(\hat{\bbox{p}}_\pm)} \mbox{sgn}\, v_x\delta
g_\pm(\hat{\bbox{p}}_\pm,x,\omega_m),\\
f_\pm^+(\hat{\bbox{p}}_\pm,x,\omega_m)=e^{-i\varphi_\pm(\hat{\bbox{p}}_\pm)}
\pm e^{-i\varphi_\pm(\hat{\bbox{p}}_\pm)}
\mbox{sgn}\, v_x\delta g_\pm(\hat{\bbox{ p}}_\pm,x,\omega_m),\\
g_\pm(\hat{\bbox{p}}_\pm,x,\omega_m)=\delta
g_\pm(\hat{\bbox{p}}_\pm,x,\omega_m).
\end{array} \right.
\label{37}
\end{equation}
Here $\varphi_\pm(\hat{\bbox {p}}_\pm)$ are the phases of the order parameters
$\Delta_\pm(\hat{\bbox{p}}_\pm)$.
Eq.(\ref{37}) satisfies the normalization condition (\ref{2}).
The ansatz (\ref{37}) is correct for $\omega_m \ll \Delta(\hat{\bbox{p}})$.
The latter inequality in turn holds for the parameter region
$v_F/d \ll T \ll |\Delta_{\pm}(\theta=0)|$ or  $T \ll v_F/d$,
$d \gg \xi_0$  which will be considered below.

Substituting (\ref{37}) into (\ref{1}) in the main approximation one obtains
 \begin{equation}
|v_x| \nabla_x \delta g_\pm (\hat{\bbox{p}},x,\omega_m)=
\mp2|\Delta_\pm(\hat{\bbox{ p}}_\pm,x)|\delta g_\pm(\hat{\bbox{p}}_\pm,x,\\
\omega_m) .
\label{38}
\end{equation}

The solution of the Eilenberger equations (\ref{1}) in the normal metal
is trivial. Combining this solution with (\ref{37}) and making use of the
continuity condition for the Green functions at NS interfaces we find
\begin{equation}
\left\{ \begin{array}{l}
g_N(\hat{\bbox{p}},\omega_m)=\delta g_-(\hat{\bbox{p}}_-,\omega_m)=
\delta g_+(\hat{\bbox{p}}_+,\omega_m),\\
(e^{i\varphi_-(\hat{\bbox{p}}_-)}+e^{i\varphi_-(\hat{\bbox{p}}_-)} \mbox{sgn}\,
v_x\delta g_-(\hat{\bbox{p}}_-,\omega_m)) e^{-2\omega_m d/v_x}=\\
=e^{i\varphi_+(\hat{\bbox{p_+}})}-e^{i\varphi_+(\hat{\bbox{p}}_+)} \mbox{sgn}\,
v_x\delta g_+(\hat{\bbox{p}}_+,\omega_m),
\end{array} \right.
\label{39}
\end{equation}
$g_N$ is the Eilenberger Green function in the normal metal.
Similarly to the case of conventional superconductors (see e.g.\cite{zaik})
the equations (\ref{39}) yield
\begin{equation}
g_N(\hat{\bbox{p}},\omega_m)=\mbox{sgn}\, v_x \tanh \{
\frac{i\varphi(\hat{\bbox{
p}})}{2}+\frac{\omega_md}{v_x}\},
\label{40}
\end{equation}
where $\varphi(\hat{\bbox{p}})=\varphi$ for
$\psi_+(\hat{\bbox{p}}_+) \psi_-(\hat{\bbox{p}}_-) >0$
and $\varphi(\hat{\bbox{p}})=\varphi+\pi$
for $\psi_+(\hat{\bbox{p}}_+) \psi_-(\hat{\bbox{p}}_-) <0$ (as before
the gap is chosen to
be proportional to $\psi_-(\hat{\bbox{p}}_-)$ on the left side and to
$\exp(i\varphi)\psi_+(\hat{\bbox{p}}_+)$ on the right side of the barrier).

Substituting (\ref{40}) into (\ref{17}) we arrive at the final expression
for the Josephson current in SNS junctions. For $v_F/d \ll T \ll
|\Delta_{\pm}(\theta=0)|$
we reproduce the standard result
$$
j_{S}=6en\exp(-2\pi Td/v_F )\sin \varphi'/md,
$$
derived before for conventional superconductors \cite{kul69,ishii}.  Here
$\Delta_{\pm}(\theta=0)=\Delta(p_{x\pm}=p_F)$ and
$\varphi'$ is the total phase difference between $\Delta_+(\theta =0)$ and
 $\Delta_-(\theta=0)$, $n=p_F^3/3\pi^2$ is the electron concentration. The
difference between s- and d-wave superconductors becomes important in the low
temperature limit $T \ll v_F/d$. At $T\rightarrow 0$ and $d \gg \xi_0$ we get
\begin{equation}
j_{S}=\frac{3en}{4\pi md} (C_1[\varphi]+C_2[\varphi+\pi]).
\label{42}
\end{equation}
The function $[\varphi]$ defines the standard sawtooth behavior of
$j_S(\varphi )$ for s-wave superconductors at $T=0$ \cite{ishii} (see Fig.4)
and
$$
C_1=\int_{v_{x}>0}\cos^2\theta d\Omega^+,\;\; C_2=\int_{v_{x}>0}\cos^2\theta
d\Omega^-,
$$
$d\Omega^{+,-}$ are the solid angle elements on the Fermi sphere for which
the functions $\psi_-(\hat{\bbox{p}}_-)$ and $\psi_+(\hat{\bbox{p}}_+)$ have
equal or opposite signs respectively. The phase dependence of the Josephson
current in SNS junctions between d-wave superconductors $j_S(\varphi)$
(\ref{42}) is presented in Fig.5a. In contrast to the analogous dependence
for s-wave superconductors (Fig.4) it contains an additional jump at $\varphi
=0$.
This jump is due to the presence of an additional phase shift $\pi$ acquired by
electrons with momentum directions corresponding to different signs of the gap
functions in two superconductors.

We believe that the above mentioned unusual behavior of d-wave SNS junctions
can be used
to provide an experimental test for the symmetry of the order parameter in high
temperature superconductors. Let us consider a superconducting
ring interrupted by an SNS junction with the current-phase relation (\ref{42}).
Rewriting this relation in the form $j_{S}=A_1\varphi \mp A_2\pi$ respectively
for $0<\varphi<\pi$ and $-\pi<\varphi<0$ one can easily derive the free energy
of the system $F$.  In the absence of an external magnetic field we have
\begin{equation}
 F(I)=\frac{L}{2}(I^2+\kappa I^2-2I(0)|I|),
\label{free}
\end{equation}
 where $I$ is the current in the ring, $L$ is the ring inductance,
 $I(0)=A_2\pi$, $\kappa=2\pi LA_1/\Phi_0$,  $\Phi_0$ is the flux
 quantum. This expression is valid for $|2LI/\Phi_0|<1$, for larger values of
 $|I|$ the two last terms in (\ref{free}) are periodically continued with the
 period $\Phi_0/L$. The free energy of a SQUID with an SNS junction is
shown in Fig.5b for the large inductance limit. It has two minima at
$\varphi = \pm \pi A_2/A_1$ which correspond to the condition
$I_S(\varphi )=0$. Thus an SNS junction between two d-wave superconductors
has a twofold degenerate ground state inside the interval $-\pi <\varphi \leq
\pi$.
This behavior differs from that for tunnel junctions in which case the system
has
only one energy minimum at $\varphi =0$ or $\varphi =\pi$.

Minimizing (\ref{free}) with respect to $I$ we find the
 equilibrium value for the current
 $$I=\pm I(0)/(1+\kappa).
 $$ This result means that an SNS junction described by the current-phase
relation (\ref{42}) $always$
 induces a spontaneous current in a superconducting ring no matter how small
 the inductance $L$ is. This result differs from that obtained for a ring with
a
 $\pi$-junction \cite{Bul} in which case the spontaneous superconducting
current
 can occur only provided $L$ is sufficiently large.

 Without an external magnetic field the ground state of the system is
degenerate
 with respect to the direction of the current $I$ flowing across the ring. This
 degeneracy is lifted by an external magnetic flux $\Phi$ applied to the ring.
 In this case the value $I$ in the last two terms of (\ref{free})
 should be substituted by $I+(\Phi/L)$ and the energies of the two lowest
states
 differ by $\Delta F=2I(0)\Phi/(1+\kappa)$. For $\kappa \gg 1$ and
 $ A_1\sim A_2$  we obtain a simple estimate $\Delta F \sim \Phi_0\Phi /L$.

If one considers a SQUID configuration with two SNS junctions  one can easily
see that
 the critical current through this system $I_{max}$ may reach its minimum value
 not only
 at $\Phi/\Phi_0=0$ (as in the case of a SQUID with 0-junctions) or at
$\Phi/\Phi_0=1/2$ (as for a SQUID with  $\pi$-junctions) but at an arbitrary
value of $\Phi/\Phi_0$ depending on the relation between $A_1$ and $A_2$.
 The dependence $I_{max}(\Phi )$ for a SQUID with identical SNS junctions
 and $A_1\ge 2A_2$ is depicted in Fig.6. The minimum value of $I_{max}$
 is reached at $\Phi/\Phi_0=(A_1-A_2)/2A_1$.

\subsection{Short Weak Links.}

In addition to tunnel junctions  and SNS structures another type of weak links
between d-wave
superconductors is of physical interest. Let us consider two superconductors
separated by an impenetrable insulating barrier with a small orifice of a
typical size $L\ll \xi_{0}$. Below we shall assume that electrons can
freely move (rather than tunnel) through this orifice and put its transparency
coefficient
equal to one $D(\hat{\bbox{ p}})=1$. This model describes various geometries
(microconstrictions, microbridges etc.) which provide a direct contact
between two metals. The normal state conductance of such
systems depends only on the cross section area of the orifice ${\cal A}$ and
is given by the well known expression for the inverse Sharvin resistance
\begin{equation}
1/R_o=e^2p_F^2 {\cal A}/4\pi^2.
\label{31}
\end{equation}
In the case of conventional superconductors the dc Josephson effect in this
type
of weak links was studied in details by Kulik and Omel'yanchuk \cite{kul}.
It was found in \onlinecite{kul} that at low temperatures the corresonding
current-phase relation deviates from the standard $\sin \varphi$-form leading
to a somewhat higher Josephson critical current than that for tunnel junctions
\cite{AB}. Here we briefly discuss the generalization of the theory \cite{kul}
for the case of d-wave superconductors.

Following \cite{kul} we shall assume that the gap function $\Delta$ is not
disturbed
in superconducting bridges due to the presence of a microconstriction. This
assumption is valid everywhere except for a narrow region $\delta r \ll
\xi_{0}$
close to the orifice. It is straightforward to check that the particular form
of
$\Delta$ in this region is not important for calculation of the current through
the orifice. Therefore without loss of generality (and also for the sake of
definiteness)
we stick to the same form of the order parameter $\Delta (x)$ in two
superconducting bridges as that discussed before for the case of tunnel
junctions.

First let us cosider crystal orientations $\psi_{\pm}(\hat{\bbox{p}})
\approx \psi_{\pm}(\check{\bbox{p}})$
for which the value $\Delta$ is (nearly) uniform in both superconductors. Then
following the procedure \cite{kul} one can easily solve the Eilenberger
equations
in superconductors. Matching the Green functions at $x=0$ for the electron
trajectories passing through the orifice and assuming these functions to be
equal
to the equilibrum ones far from the weak link $x \rightarrow \pm \infty$
similarly to
\cite{kul} we obtain the expression for the superconducting current through
the orifice $I_S=j_S{\cal A}$
\begin{equation}
I_S=8\pi eT{\cal A}\sum_{m>0}\int_{v_{x}>0}
\frac{v_{x}(\hat{\bbox{p}}_-)
  \Delta_+(\hat{\bbox{p}}_+)\Delta_-(\hat{\bbox{p}}_-)\sin\varphi }{\omega_m^2+
[(\omega_m^2+\Delta_+^2(\hat{\bbox{p}}_+))
(\omega_m^2+\Delta_-^2(\hat{\bbox{p}}_-))]^{1/2}+
\Delta_+(\hat{\bbox{p}}_+)\Delta_-(\hat{\bbox{p}}_-)\cos\varphi}
\frac{d^2S_-}{(2\pi)^3v_F}
\label{33}
\end{equation}
At $T \gg \Delta$ Eq.(\ref{33})  reduces to Eq.(\ref{21}) with
$D(\hat{\bbox{p}})=1$. To analyse the result (\ref{33}) at lower temperatures
it is again convenient to introduce the quantity
$\varphi(\hat{\bbox{p}})$.
Then for $|\Delta_+(\hat{\bbox{p}}_{+})|\approx |\Delta_-(\hat{\bbox{p}}_{-})|$
similarly to the case of conventional superconductors \cite{kul} from
(\ref{33}) we get \begin{equation} I_{S}=2\pi e \int_{v_{x}>0}
|\Delta_-(\hat{\bbox{p}}_{-})| \sin[\varphi(\hat{\bbox{p}})/2] \tanh\biggl(
\frac{|\Delta_-(\hat{\bbox{p}}_{-})|
\cos[\varphi(\hat{\bbox{p}})/2]}{2T}\biggr)
v_{x}(\hat{\bbox{p}}_{-})\frac{d^{2}S'}{(2\pi^{3})v_{F}}   .
\label{34}
\end{equation}
Here  $d^2S' $ is the element of the Fermi sphere  for which the condition
$|\Delta_+ (\hat{\bbox{p}}_{+})|
\approx |\Delta_- (\hat{\bbox{p}}_{-})|$ is satisfied.
As in \cite{kul} the current turns out to be discontinuous at $\varphi = \pi$.
In the
opposite limit $|\Delta_+(\hat{\bbox{p}}_{+})|\gg
|\Delta_-(\hat{\bbox{p}}_{-})|$ and at $T=0$  with the logarithmic accuracy we
find ($0\le\varphi(\hat{\bbox{p}})\le2\pi$) $$ I_{S}=4e \int_{v_{x}>0}
\sin\varphi(\hat{\bbox{p}}) |\Delta_-(\hat{\bbox{p}}_{-})| \ln
\frac{|\Delta_+(\hat{\bbox{ p}}_{+})|}{|\Delta_-(\hat{\bbox{p}}_{-})|}
v_{x}(\hat{\bbox{ p}}_{-})\frac{d^{2}S^{''}}{(2\pi)^{3}v_{F}}, \phantom{dd}
\mbox{
for} \quad |\varphi(\hat{\bbox{p}})-\pi|\gg \ln^{-1}
\frac{|\Delta_+(\hat{\bbox{p}}_{+})|}{|\Delta_-(\hat{\bbox{p}}_{-})|}, $$
\begin{equation} I_{S}= -4\pi e \int_{v_{x}>0}\mbox{sgn}\,
(\varphi(\hat{\bbox{p}})-\pi) |\Delta(\hat{\bbox{ p}}_{-})|
v_{x}(\hat{\bbox{p}}_{-})\frac{d^{2}S^{''}}{(2\pi)^{3}v_{F}}, \phantom{dd}
\mbox{for}
\quad |\varphi(\hat{\bbox{p}})-\pi|\ll \ln^{-1} \frac{|\Delta(\hat{\bbox{
p}}_{+})|}{|\Delta(\hat{\bbox{p}}_{-})|},
\label{35}
\end{equation}
where $d^2S^{''}$ denotes the element of the Fermi sphere with
$|\Delta_+(\hat{\bbox{p}}_{+})|
\gg |\Delta_-(\hat{\bbox{p}}_-)|$.
We see that the magnitude of the current jump at $\varphi(\hat{\bbox{p}})=\pi$
(\ref{35})
is by the factor $\sim \ln^{-1}(|\Delta_{+}|/|\Delta_{-}|)$  smaller as
compared
to the case $|\Delta_{+}|=|\Delta_{-}|$ (\ref{34}).
For $\varphi(\hat{\bbox{p}})$ close to $\pi$ and arbitrary ratio
  $(|\Delta_{+}|/|\Delta_{-}|)$ the magnitude of the jump reads
\begin{equation}
I_{S}= -4\pi e
\int_{v_{x}>0}^{'}\mbox{sgn}\, (\varphi(\hat{\bbox{p}})-\pi)
\frac{|\Delta(\hat{\bbox{p}}_{+})|
|\Delta(\hat{\bbox{p}}_{-})|}{|\Delta(\hat{\bbox{p}}_{+})|+|\Delta(\hat{\bbox{
p}}_{-})|} v_{x}(\hat{\bbox{p}}_{-})\frac{d^{2}S}{(2\pi)^{3}v_{F}}.
\label{36}
\end{equation}
The integration in (\ref{36}) runs over the parts of
Fermi-surface where $\varphi(\hat{\bbox{p}})$ is close to $\pi$. As the
function
 $\psi(\hat{\bbox{p}})$ changes its sign on the Fermi-surface an additional
  jump on the $I_S(\varphi)$ dependence takes place at $T\rightarrow 0$
  similarly to the case of SNS junctions.

Let us emphasize again that the result (\ref{33}) holds only for a
homogeneous distribution of the order parameter in superconducting banks.
Within the same framework an analogous result was recently derived by
Yip \cite{yip}. Provided the condition $|\Delta(\hat{\bbox{p}})|\approx
|\Delta(\check{\bbox{p}})|$ is not satisfied the superconducting
order parameter depends on the coordinate and the expression for Josephson
current deviates from (\ref{33}).  In this case after a straightforward
calculation
one can show that at
$T \gg \Delta (T)$  the value $I_S$ is given by Eqs.
(\ref{22}), (\ref{23}) and hence again reduces to the result (\ref{29}) with
$D(\hat{\bbox{p}})=1$ ($k=0$) and $R_N \rightarrow R_o$.

\section{Quasiparticle Tunneling and Phase Fluctuations}

\subsection{Low Voltage Conductance and I-V Curve}

A possible way to test the symmetry of the superconducting order parameter
is to measure the I-V curve of a tunnel junction in the limit of low
temperature
and voltage. In the case of isotropic s-wave superconductors at $T \ll \Delta$
only a small number of quasiparticles activated above the gap contributes to
the junction
conductance $G$. Therefore in the limit of small voltages we have $G \propto
\exp (-\Delta /T)$.
At $T=0$ no quasiparticles exist above the gap and the current across the
junction is equal to zero $I=0$ provided the externally applied voltage $V$
does not exceed the value $\Delta /e$ for NS junctions and $2\Delta /e$ for SS
junctions. Below we shall show that in the case of d-wave symmetry of the order
parameter the I-V curve of a tunnel junction is entirely different in the
corresponding temperature and voltage intervals.

Let us assume that the time-independent external voltage $V$ is applied to the
tunnel junction between two metals. Then expressing the current in terms of the
Green function of the system and making use of the boundary conditions
(\ref{4})
in the lowest order in $D$ after a standard calculation (see e.g. \cite{zai})
one easily finds
\begin{equation}
j=e\int\biggl(\int_{v_x>0}\biggl{[}\tanh \biggl{(}\frac{\epsilon}{2T}\biggr{)}-
\tanh \biggl{(}\frac{\epsilon-eV}{2T}\biggr{)}\biggr{]}
g'_+(\epsilon-eV,\hat{\bbox{p}}_+)
g'_-(\epsilon,\hat{\bbox{p}}_-)d\epsilon\biggr) v_x(\hat{\bbox{p}}_-)
D(\hat{\bbox{p}}_-) \frac{d^2S_-}{(2\pi)^3v_F}.
\label{disscur}
  \end{equation}
Here $j_N$ is a dissipative contribution to the current across the junction and
$g'_{\pm}(\epsilon,\hat{\bbox{p}}_\pm )$ are the normalized densities of states
  of two metals in the vicinity of a tunnel barrier. In the case of d-wave
superconductors for $\Delta(\hat{\bbox{p}}) =\Delta(\check{\bbox{p}})$ we have
\begin{equation}                      g'_{\pm}(\epsilon,\hat{\bbox{p}}_\pm )=
\frac{|\epsilon |\Theta (|\epsilon |-|\Delta _{\pm}(\bbox{p}_{\pm})|)}
{\sqrt{\epsilon^2-\Delta^2_{\pm}(\bbox{p}_{\pm})}}.
\label{density}
\end{equation}

Let us first calculate the I-V curve of an NS junction. Setting $\Delta_-=0$
and
$\Delta_+ =\Delta(\hat{\bbox{p}})$ and substituting (\ref{density}) into
(\ref{disscur}) we obtain at $T=0$
\begin{equation}
 j_N=2e  \int_{v_{x}>0}
 [(eV)^2-\Delta^2(\hat{\bbox{p}})]^{1/2}
 \Theta (eV-|\Delta (\hat{\bbox{p}})|)
          D(\hat{\bbox{p}}) v_{x}(\hat{\bbox{p}})\frac{d^{2}S}{(2\pi)^{3}
v_{F}}.
\label{44}
\end{equation}
The equation (\ref{44}) defines the dissipative current across the tunnel
junction for crystal orientations with $\Delta(\hat{\bbox{p}}) \approx
\Delta(\check{\bbox{p}})$. For other crystal orientations the superconducting
density of states in the vicinity of a tunnel barrier deviates from
(\ref{density}).  Nevertheless -- as in the case of a dc Josephson current --
at $T=0$ the result (\ref{44}) remains to be valid apart from an unimportant
numerical factor of order one. Below we shall neglect this factor and apply the
result (\ref{44}) to any crystal orientation. Then choosing the order
parameter in the form $\Delta (\hat{\bbox{p}})=\Delta_0(p_{x_0}^2-p_{y_0}^2)$
from Eq. (\ref{44}) we have $$
j_N=\frac{ep_F^2}{4\pi^3}\int_{0}^{2\pi}d\phi\int_{0}^{\pi/2}d\theta \sin\theta
\cos\theta D(\theta) \{(eV)^2-\Delta_0^2[\sin^2\theta \cos^2\phi
$$
\begin{equation}
- (\sin\theta \sin\phi \cos\theta_0-\cos\theta
\sin\theta_0)^2]^2\}^{1/2} .
\label{45}
\end{equation}
Here $\theta_0$ is the angle between the vector $\bbox{ n}$ normal to the
junction plane
and the crystal axis $z_0$, $\Delta_0$ is the maximal value of
$\Delta(\hat{\bbox{p}})$.

It is easy to see that -- in contrast to the case of s-wave superconductors --
the current (\ref{45}) does not vanish  even for $eV\ll \Delta_0$. In the
latter
limit the main contribution to $j_N$ comes from quasiparticles with the
momentum directions close to the directions for which the order
parameter $\Delta (\hat{\bbox{p}})$ is equal to zero.  The integral over these
momentum directions can be in turn splitted into two terms \begin{equation}
j_N=j_{N1}+j_{N2}.
\label{twoterms}
\end{equation}
The first term $j_{N1}$ is defined by the integral
over the values $\theta$ close to $\theta_0$ and all values of $\phi$ from
$0$ to $2\pi$ or, in other words, over the momentum values
$p_{z_0} \approx p_F$. With the logarithmic accuracy the corresponding
integration in (\ref{45}) yields
\begin{equation}
j_{N1}=\frac{ev_{x}(\theta_0)D(\theta_0) (eV)^2p_F^2}{8\pi^2 v_F\Delta_0}\ln
\frac{\Delta_0}{eV}.
\label{46}
\end{equation}
The second term $j_{N2}$ comes
from the integration over momentum directions close to the lines $p_{x_0}=\pm
p_{y_0}$. In the vicinity of these lines the gap function is
$\Delta(\hat{\bbox{p}})=(p_2/p_F)\Delta_0 h(p_1)$ where $p_1$ is the coordinate
along the line of zeros and $p_2$ the one in perpendicular direction. In our
case $h(p_1)=2\sin \theta'$, where $\theta'$ is the angle between
$\hat{\bbox{p}}$ and $z_0$. Integrating over $p_2$ we obtain
\begin{equation}
j_{N2}=e\biggl[ \int_{v_{x}>0}
\frac{D(p_1)v_{x}(p_1)}{|h(p_1)|} dp_1\biggr]\frac{(eV)^2 p_F}{8\pi^2\Delta_0
v_F}.
\label{48}
\end{equation}
Comparing the results (\ref{46}) and (\ref{48})
one can conclude that in the limit $eV \ll \Delta_0$ the current $j_{N1}$
dominates for crystal orientations
$\pi/2-|\theta_0| \gg \ln^{-1/(k+1)}(\Delta_0/eV)$. E.g.  for $\theta_0=0$ and
$D(\theta)\propto
\cos^2 \theta$ ($k=2$) we get
\begin{equation}
j_{N1}=\frac{eV^2}{R_N\Delta_0}  \ln \frac{\Delta_0}{eV}.
\label{47}
\end{equation}
For $\pi/2-|\theta_0| \ll
\ln^{-1/3}(\Delta_0/eV)$ the axis $z_0$ nearly coincides with the junction
plane and the term $j_{N1}$ becomes small. In this case the current $j_N$ is
given by the term $j_{N2}$ (\ref{48}).  For $\theta_0=\pm \pi /2$ and $eV\ll
\Delta_0$ it yields
\begin{equation}
j_{N2}=\frac{\pi}{4\sqrt{2}}\frac{eV^2}{R_N\Delta_0}.
\label{49}
\end{equation}

The zero temperature I-V curves for an NS tunnel junction with
$D \propto p_x^2/p_F^2$ are presented in Fig.7 for two particular crystal
orientations
(one of the principal axes $x_0$ or $z_0$ is perpendicular to the barrier
plane).
At low voltages $eV \ll \Delta_0$ these curves follow the results (\ref{47}),
(\ref{49})
(improving the logarithmic accuracy of (\ref{47}): $\ln (\Delta_0/eV)
\rightarrow
\ln (2.4\Delta_0/eV))$. At higher voltages $eV \sim \Delta_0$ the I-V curves
shows
a smooth crossover to the standard Ohmic behavior.

The I-V curve of a tunnel junction between two d-wave superconductors can
be calculated
analogously. Substituting (\ref{density}) into (\ref{disscur}) for the case of
identical superconductors at $T=0$ we find
\begin{equation}
j_N=2e\int_{v_{x}>0} \biggl(\int_{|\Delta(\hat{\bbox{p}}_-)|}^{ eV-
|\Delta(\hat{\bbox{p}}_+)|} \frac{\omega(eV-\omega)d\omega}{
(\omega^2-\Delta^2(\hat{\bbox{p}}_-))^{1/2}
((eV-\omega)^2-\Delta^2(\hat{\bbox{p}}_+))^{1/2}}  \biggr)
  D(\hat{\bbox{p}}_-) v_{x}(\hat{\bbox{p}}_-)\frac{d^{2}S_-}{(2\pi)^{3}v_{F}}.
\label{50}
\end{equation}
The integration in (\ref{50}) is made over that
parts of Fermi-surface where
$eV-|\Delta_+(\hat{\bbox{p}}_+)|>|\Delta_-(\hat{\bbox{ p}}_-)|$. In a general
case the zero lines of the order parameters in two superconductors do not
coincide. Assuming that the angle $\chi$ between these lines in the point
of their intersection $\hat{\bbox{p}}_i$ obeys the condition
$eV/\Delta_0|h(\hat{\bbox{p}}_i)|\chi\ll 1$, we can easily obtain the leading
order contribution
to the quasiparticle current for $eV\ll \Delta_0$:
\begin{equation}
j_N=\frac{ev_{x}(\hat{\bbox{p}}_i) D(\hat{\bbox{p}}_i) p_F^2 (eV)^3}{24\pi v_F
|h_+(\hat{\bbox{p}}_{i+})   h_-(\hat{\bbox{p}}_{i-})|
|\sin\chi|\Delta_{0}^2}.
\label{52}
\end{equation}
In order to find the
total current it is necessary to sum up the contributions from all intersection
points.
Then one obtains
\begin{equation}
j_N=a\frac{e^2V^3}{R_N\Delta_0^2}.
\label{jN}
\end{equation}
Here the factor $a$ keeps track on the particular relative orientation of
superconductors and is of order one for most of such orientations. This
factor vanishes only provided $v_x (\hat{\bbox{p}_i})
D(\hat{\bbox{p}_i}) = 0$, i.e. if the intersection points coincide with the
poles of the Fermi surface and the $z_0$ axes of both superconductors are in
the junction plane.

If both superconductors are oriented identically we again arrive with
the aid of (\ref{50}) at the expression for the current $j_N$ defined by the
expressions  (\ref{46})-(\ref{49})
multiplied by the numerical prefactor
$$
\frac{8}{\pi} \int_{0}^{1/2} \omega K\biggl( \frac{\omega}{1-\omega}\biggr)
d\omega   \approx 0.6,
$$
where as before $K(t)$ is the complete elliptic
integral. In this case the zero lines  of
$\Delta_-(\hat{\bbox{p}}_-)$, $\Delta_+(\hat{\bbox{p}}_+)$ coincide if they are
drawn on the same Fermi-surface.

The zero temperature I-V curves for the junction between two
d-wave superconductors calculated numerically from the equation
(\ref{50}) are presented in Fig.8. The upper and the lower curves
were calculated assuming that
respectively
$z_0$ and $x_0$ axes of both superconductors are perpendicular to the boundary
plane. Orientation of $x_0$ and $y_0$ axes of two superconductors is identical
for the upper curve whereas for the lower curve their $y_0$ axes constitute
the angle $\pi /2$ between each other. Again in the low voltage limit the
numerical curves agree well with the analytic results (\ref{jN}), (\ref{47})
and allow to define the corresponding numerical prefactors. E.g. the factor $a$
in Eq. (\ref{jN}) is found to be equal to $a \sim 0.35$ for the above crystal
orientation
and the logarithmic accuracy of Eq. (\ref{47}) can be improved by a
substitution
$\ln (\Delta_0 /eV) \rightarrow \ln (3.9\Delta_0 /eV)$.

For larger voltages of order $\Delta_0$ the differential conductance $G=dI/dV$
has a maximum which exact position depends on the relative crystal orientation.
E.g. for the two curves presented in Fig.8 the conductance $G(V)$ reaches its
maximum respectively at $eV \simeq 1.05\Delta_0$ and
$eV \simeq 1.97\Delta_0$.

In order to understand the difference in the maximum positions by nearly a
factor
2 for these two configurations let us consider the case
$\Delta_+(\hat{\bbox{p}}_+)\approx \Delta_-(\hat{\bbox{ p}}_-)$ and carry out
the frequency integration in Eq.(\ref{50}). Then we obtain
\begin{equation}
\frac{dI}{dV}R_N=\frac{
\int_{v_x>0}k^{-2}[E(k)(1+k^2)-(1-k^2)K(k)]
D(\hat{\bbox{p}}_-)v_x(\hat{\bbox{p}}_-) d^2S_-}{2\int_{v_x>0}
D(\hat{\bbox{p}}_-)v_x(\hat{\bbox{p}}_-) d^2S_-},
\label{div}
\end{equation}
where $k(\hat{\bbox{p}})=((eV)^2-(2\Delta(\hat{\bbox{p}}))^2)^{1/2}/eV$ and
$E(k)=\int_{0}^{\pi/2}(1-k^2\sin^2\phi)^{1/2}d\phi$.  The momentum integration
in (\ref{div}) is carried out over that parts of Fermi surface where $eV>
2|\Delta(\hat{\bbox{p}})|$.
It is easy to see that the maximum of $dI/dV$ takes place at a certain
effective gap value which (due to the presence of the factor
$D(\hat{\bbox{p}}_-)v_x(\hat{\bbox{p}}_-)$ in the integrand (\ref{div}))
is mainly defined by the value of gap function in the direction normal to
the junction plane. Since for the lower
curve of Fig.8 for both superconductors the gap function is equal to
$\Delta_0$ in this direction the maximum of $dI/dV$ takes place at
$eV \approx 2\Delta_0$. For the upper curve the gap vanishes along the
direction
normal to the interface. Accordingly the maximum has a much smaller amplitude
and takes place at lower voltages $eV \approx \Delta_0$.

Note that the behavior of the low voltage conductance $G \propto V^2$
has been detected in recent experiments with SS tunnel junctions \cite{Mand}.
This behavior is in a good agreement with our theoretical predictions
(\ref{52})
and (\ref{jN}). Also for higher voltages our results qualitatively agree with
those reported in \cite{Mand}.

The dependence $G \propto V^2$ for the low voltage conductance of a tunnel
junction between d-wave superconductors has been also discussed in a recent
paper
by Won and Maki \cite{WM} within a different theoretical framework. They
evaluated
the quasiparticle current by means of the standard tunneling Hamiltonian
approach
assuming that tunneling matrix elements are independent of the momenta of
tunneling electrons and making use of the expressions for the superconducting
densities of states averaged over all momentum directions. This approach yields
the results which are independent of relative orientation of two
superconductors.
Although it appears to be quite difficult to justify such an approach
microscopically
we believe that it might work -- at least qualitatively -- for
diffusive SS boundaries.
However it clearly fails for specularly reflecting boundaries in which case the
quasiparticle
current essentially depends on the relative orientation of d-wave
superconductors.

Very recently the case of specularly reflecting boundaries has been
independently
studied by Bruder, van Otterlo and Zimanyi \cite{BOZ}. These authors also
proceeded
within the tunneling Hamiltonian approach completed by a phenomenological
assumption about the angular dependence of the tunneling matrix elements. For
identically
oriented superconductors with $z_0$ axes being in the barrier plane they also
arrived at the result $j_N \propto V^2$ which agrees with our results
(\ref{48}),
(\ref{49}). However for misoriented superconductors at $T=0$ a vanishing subgap
current
$I(eV \ll \Delta_0)=0$ has been found in \cite{BOZ}. In contrast our results
(\ref{46}), (\ref{48}) and (\ref{52}) demonstrate that even at $T=0$ the subgap
current does not vanish for all crystal orientations with $v_x
(\hat{\bbox{p}_i})
D(\hat{\bbox{p}_i}) \neq 0$ \cite{FN3}, $\hat{\bbox{p}_i}$ is value of the
electron momentum at the intersection point of the nodal lines for the two
order parameters $\Delta_+$ and $\Delta_-$.
The origin of this disagreement lies in the fact that the authors \cite{BOZ}
considered the case of a long cylindrical Fermi surface whereas
the analysis developed here is based on a picture of a (nearly) spherical
Fermi surface. As under certain restrictions the Eilenberger formalism can be
also applied to superconductors
with nonspherical Fermi surfaces \cite{Al} the corresponding generalization of
our approach can be easily provided. E.g. Eqs.(\ref{RN}), (\ref{52}) remain
valid
 for the metals described by  the dispersion law
$\epsilon (\bbox{p})=(p_{x_0}^2+p_{y_0}^2)/2m_1 + p_{z_0}^2 /2m_2 $.
In order to find the expressions for $h_+$ and $\chi $ in (\ref{52}) one should
rewrite the function
$\Delta_+(\hat{\bbox{p}}_+)$ in terms of the variable
$\hat{\bbox{p}}_-$ and draw the nodal line of this function on the
Fermi-surface
of the ``-''-metal.  Then it is easy to see that the prefactor $a$ in
Eq.(\ref{jN})
turns out to be of order
$a \sim (m_1/m_2)^{1/2}$.  It appears that the results of \onlinecite{BOZ}
correspond
to the limiting case $m_1/m_2 \rightarrow 0$ for which the prefactor $a$ in
(\ref{jN})
(and thus the current $j_N$) vanishes at $T=0$ and small $V$. Physically this
situation corresponds to ideally two-dimensional character of the electron
motion
in $x_0-y_0$ planes whereas in the $z_0$-direction this motion is totally
suppressed.
If one allows for jumps of electrons  between such planes the ratio $m_1/m_2$
differs from zero and the result $j_N \propto V^3$ remains valid. It is easy to
check
that the validity condition for this result $\chi \gg eV/\Delta_0$ remains
unchanged for any nonzero $m_1/m_2$ whereas the result $j_N \sim V^2$ holds
only in the
narrow region of misorientations of two superconductors $\chi
\ll(eV/\Delta_0)(m_1/m_2)^{1/2} $.

AT low but finite temperatures $T \ll \Delta_0$ the results obtained above
for the case $T=0$ remain valid for not very small voltages $eV \gg T$.
In the opposite limit $eV \ll T$ the main contribution to the current $j_N$
comes from quasiparticles thermally activated above the gap. In the limit
$eV \ll T \ll \Delta_0$ for the NS junction we find
\begin{equation}
j_N=G(T)V,
\label{SNT}
\end{equation}
where the linear conductance $G(T)$ of a tunnel junction between misoriented
superconductors is
\begin{equation}
G(T) \sim T^2/R_N\Delta_0^2.
\end{equation}
Analogously for the crystal orientations described by the equations (\ref{46})
and
(\ref{48}) one gets respectively $G(T) \sim (T/R_N \Delta_0)\ln (\Delta_0/T)$
and
$G(T) \sim (T/R_N \Delta_0)$.

\subsection{Effective Action}
Finally let us briefly demonstrate how the above results can be generalized
to take into account thermodynamic and quantum fluctuations of the phase
difference
$\varphi$ across the tunnel junction between d-wave superconductors. The grand
partition
function of this junction can be expressed in terms of the path integral
over the $\varphi$-variable (see e.g. \cite{SZ})
\begin{equation}
Z \sim \int {\cal D} \varphi (\tau ) \exp (-S_{eff}[\varphi (\tau )]),
\label{partfunc}
\end{equation}
$\tau$ is the imaginary time variable which changes from $0$ to $\beta =1/T$.
To evaluate the effective action functional $S_{eff}[\varphi]$
we make use of the approach developed in Refs. \cite{ZP85,SZ} which allows to
recover $S_{eff}[\varphi]$ from the expression for the kernel of the current
density
operator $\bbox{j}(\bbox{r}, \tau)$ by means of the integration over the
effective
``coupling constant'' $\lambda$. In our case the corresponding formula reads
\begin{equation}
S[\varphi]=\int^1_0 d\lambda \int^{\beta}_0 d\tau \varphi (\tau)
j[\lambda \varphi (\tau)] {\cal A}/2e,
\label{trick}
\end{equation}
where $j[\varphi (\tau)]$ represents the current density through the junction.
In the interesting limit of low frequencies the expression for the supercurrent
$j_S$ reduces to the standard Josephson relation
\begin{equation}
j_S=j_0\sin \varphi (\tau ),
\label{qj}
\end{equation}
whereas the kernel for the quasiparticle current operator has the form
\cite{SZ}
\begin{equation}
j_N[\varphi (\tau )]=2e \int_0^{\beta} d\tau '\alpha (\tau - \tau ')
\sin \biggl{(} \frac{\varphi(\tau)-\varphi(\tau')}{2} \biggr{)}.
\label{qq}
\end{equation}
Combining (\ref{trick})-(\ref{qq}) and
also taking into account the charging energy term one immediately arrives at
the
AES effective action \cite{AES}
\begin{equation}
S_{eff}= \int_0^{\beta} d\tau \Big{[}\frac{C}{2}
\Big{(}\frac{\dot{\varphi}}{2e}\Big{)}^2 - E_J \cos \varphi (\tau )\Big{]}
-\int_0^{\beta}d\tau \int_0^{\beta}d\tau'
\alpha (\tau-\tau') \cos
\biggl{(}\frac{\varphi(\tau)-\varphi(\tau')}{2}\biggr{)},
\label{action}
\end{equation}
C is the junction capacitance and $E_J=j_0{\cal A}/2e$ is the Josephson
coupling energy
which can be positive or negative depending on the relative crystal
orientation.
The particular form of $\alpha (\tau )$ depends on the form of the I-V curve in
the limit of small $V$. E.g. making use of (\ref{qq}) it is easy to show that
for $I \propto V^3$  one has $\alpha (\tau ) \propto \tau ^{-4}$. More
precisely,
combining (\ref{jN}) with (\ref{qq}) we obtain at $T \rightarrow 0$
\begin{equation}
\alpha (\tau )=3a/\pi e^2R_N\Delta_0^2\tau^4.
\label{alpha}
\end{equation}
According to the above analysis this result holds for most of crystal
orientations.
For the orientations describrd by the I-V curves (\ref{47}) and (\ref{49})
we find respectively $\alpha (\tau ) \sim \ln (\Delta_0 \tau )/e^2R_N \Delta_0
\tau^3$
and $\alpha (\tau ) \sim 1/e^2R_N \Delta_0 \tau^3$. The latter dependence has
been also
obtained in \cite{BOZ}.

We believe that the above results
might be helpful for a quantitative description of thermodynamic and quantum
properties
of Josephson junctions and granular arrays composed by d-wave superconductors.

\section{Discussion}

The microscopic analysis of the charge transport in tunnel junctions and weak
links
formed by d-wave superconductors allows to encover several interesting features
of such systems. We demonstrated that the order parameter of a d-wave
superconductor
can be essentially suppressed in the vicinity of the insulating boundary
depending
on its orientation relative to the principal crystal axes of such a
superconductor.
This proximity effect can in turn strongly influence the Josephspn current
between two
superconductors and becomes particularly important at $T$ close to $T_c$. In
the
latter case the temperature dependence of the Josephson critical current $j_0$
varies from $j_s \propto T_c-T$ for a homogeneous order parameter in
superconducting
bulks (i.e. if one of the principal crystal axes is nearly perpendicular to the
junction plane) to $j_0 \propto (T_c-T)^2$ for other crystal orientations. The
results of our
calculation show a significant dependence of the Josephson current on the
relative
orientation of the superconductors and are consistent with those of recent
experiments
\cite{W,Ott,Kirtley} which indicate the possibility of d-wave pairing symmetry
in HTSC compounds.

The current-phase relation for SNS junctions and short superconducting weak
links
essentially deviates from the standard Josephson relation $j_S=j_0\sin \varphi
$
in the low temperature limit. In the case of d-wave symmetry of the order
parameter at $T=0$
the current-phase relation $j_S (\varphi )$ for SNS junctions shows an
additional jump
(as compared to the case of s-wave pairing) at the point $\varphi =0$.
Accordingly
the superconducting coupling energy for such junctions $E(\varphi )$ (in
contrast
to tunnel junctions) has two degenerate minima within the phase interval
$-\pi < \varphi \leq \pi$ (see Fig.4b) which correspond to two $different$
stable
zero current states. Positions of these minima do not coincide with
$\varphi =0$ or $\varphi =\pi$ (as for tunnel junctions) but can be located at
any point inside the interval $-\pi < \varphi \leq \pi$. Thus in the analogy
with
$0$- and $\pi$-junctions one can say that the systems in question provide an
interesting example of ``whatever-junction''. Being included into a SQUID ring
such a junction induces a spontaneous superconducting current no matter how
small
the ring inductance is and yields to further features different from those for
SQUIDs with tunnel junctions. We believe that these features could be used as
an
additional test for the pairing symmetry in HTSC.

An additional information about the form of the superconducting order parameter
and the density of states is contained in the expression for the quasiparticle
tunneling current. We evaluated this current for tunnel junctions between a
normal
metal and a d-wave superconductor as well as between two d-wave superconductors
in the low temperature limit. The corresponding I-V curves show zero-bias
anomalies
of the type $j_N \propto V^2$, $j_N \propto V^2\ln (1/V)$ or $j_N \propto V^3$
depending on the junction type and relative crystal orientations. The latter
dependence agrees well with the experimental results \cite{Mand}. At larger
voltages
the differential conductance of SS junctions has a peak (also detected
experimentally
\cite{Mand,Ved}) which position also depends on the relative crystal
orientation.

We would like to point out that one can in principle provide an example of a
$\pi$-junction
not only between d-wave superconductors but also between s-wave superconductors
with multisheet Fermi surfaces and different signs of the order parameter on
different
sheets \cite{GMG}. From this point of view an experimental confirmation of
$\pi$-junction-like
properties of HTSC compounds yet cannot completely exclude s-wave pairing. On
the other
hand, for special types of Fermi surfaces the anisotropic s-wave order
parameter
can be equal to zero for certain momentum directions not only due to the
symmetry
reasons (as it would be in the d-wave case). Therefore the low temperature
measurements of a quasiparticle tunneling current rather can be considered as a
``gaplessness'' test than really distinguish between s- and d-wave
types of pairing.
Bearing all that in mind one can conclude that it is quite important to combine
both dc Josephson effect and quasiparticle tunneling measurements for the same
tunnel junctions. Although even demonstration of combined $\pi$-junction-like
and gapless
properties of such systems  formally cannot yet exclude other than d-wave
types of pairing it would strongly favour the possibility of d-wave pairing in
HTSC.

The results derived here are not specific for HTSC compounds and can be also
applied
to other types of unconventional superconductors, like heavy fermion
superconductors.
Our analysis holds for an arbitrary form of the Fermi surface. For the sake of
definiteness some limiting results were derived for the case of a spherical
Fermi
surface. The latter is by no means restrictive for any of the concusions
reached
in the present paper. E.g. our results also remain valid for the case of a
(nearly)
cylindrical Fermi surface which appears to be more relevant for several HTSC
compounds.
The modification of our results for the latter case reduces to an effective
renormalization of the junction normal state resistance.

It is important to emphasize that our analysis is completely based on the
microscopic
theory and does not involve model assumptions which are inevitably present in
the
tunneling Hamiltonian approach. Within the latter approach the correct
dependence
of the current on the momentum directions and crystal orientations (important
for
d-wave superconductors) usually cannot be recovered in a unique way and the
validity
of the final results rather depends on physical intuition of the authors than
it is
controlled by the method itself.

After this work has been already completed we became aware of the paper
\cite{bs}
where the Josephson current between superconductors with mixed s+d symmetry of
the order
parameter has been analysed
within the tunneling Hamiltonian approach. The results obtained in this paper
differ from ours. The source of this discrepancy lies in an incorrect
phenomenological
expression for the superconducting current across the Josephson junction used
in \cite{bs}. This
expression does not account for a proper angular dependence of tunneling matrix
elements and therefore leads to irrelevant results.

We would like to thank C.Bruder, O.V.Dolgov, A. van Otterlo and G.T.Zimanyi for
useful
discussions and/or correspondence. The research described in this publication
was made possible in part (for
Yu.S.B. and A.V.G.) by  Grant No M2D000 from the International Science
Foundation. One of us
(A.D.Z) was partially supported by INTAS under Grant No 93-790, by the Russian
Foundation for Fundamental Research under Grant No 93-02-14052 and by the
Deutsche
Forschungsgemeinschaft through SFB 195.

\begin{figure}
\caption{The relative orientation of the principal crystal axes $x_0$, $y_0$,
 $z_0$ and the vector $\bbox{n}$ normal to the boundary plane.}
\label{fig.1}
\caption{The parameter $q$ (in units of $10^4v_F/T_c$) as a function of crystal
orientation relative to the boundary plane.}
\label{fig.2}
\caption{An example of the circuit which contains the odd number of
$\pi$-junctions.}
\label{fig.3}
\caption{The sawtooth function [$\varphi$] of Eq.(41).}
\label{fig.4}
\caption{The current-phase dependence (41) (a) and the energy (b)
of an SNS junction between d-wave superconductors at $T=0$.}
\label{fig.5}
\caption{The maximum current $I_{max}$ through the SQUID with two SNS junctions
(described by the current-phase dependence (41)) as a function of the
external magnetic flux $\Phi$.}
\label{fig.6}
\caption{The I-V curves for a tunnel junction between a normal metal and a
d-wave
superconductor at $T=0$. The results are for the crystal
 orientations the axis $z_0$ is perpendicular to the junction plane (upper
 solid curve) and coincides with this plane (lower solid curve). Ohmic I-V
curve
 is shown by a dashed line.}
\label{fig.7}
\caption{The same curves for a tunnel junction between two d-wave
superconductors.
The upper solid curve corresponds to similarly oriented superconductors with
their axes
$z_0$ being perpendicular to the junction plane. For the lower solid curve the
$x_0$ axes of both superconductors were taken to be perpendicular to the
junction plane
whereas their $y_0$ axes were rotated by the angle $\pi /2$ with respect to
each other.}
\label{fig.8}
\end{figure}
\newpage
\begin{picture}(100,100)
\put(50,50){\vector(1,0){40}}
\put(50,50){\vector(0,1){40}}
\put(50,50){\vector(-1,-1){23}}
\put(50,50){\vector(2,3){20}}
\multiput(70,80)(0,-6){7}{\line(0,-1){4}}
\multiput(50,50)(5,-2.5){4}{\line(2,-1){4}}
\put(90,47){$y_0$}
\put(53,90){$z_0$}
\put(30,27){$x_0$}
\put(73,80){$\bbox{n}$}
\put(50,43){$\phi_0$}
\put(51,60){$\theta_0$}
\end{picture}
\vspace{5cm}
\begin{center}
Fig.1
\end{center}
\newpage

\begin{picture}(100,100)
\put(5,5){\line(1,0){90}}
\put(5,5){\line(2,3){40}}
\put(45,65){\line(5,-6){50}}
\put(35,20){\line(1,0){48}}
\put(35,20){\line(2,3){10}}
\put(45,35){\line(5,-6){12.5}}
\put(35,20){\line(-1,0){20}}
\put(45,35){\line(-1,0){20}}
\put(47,7){\vector(0,1){8}}
\put(49,14){\large\bf z}
\put(55,12){\large 1}
\put(27,23){\vector(0,1){8}}
\put(29,30){\large\bf z}
\put(32,23){\large 2}
\put(45,38){\vector(0,1){8}}
\put(47,45){\large\bf z}
\put(51,40){\large 3}
\end{picture}
\vspace{5cm}
\begin{center}
Fig.3
\end{center}
\newpage
\begin{picture}(120,80)
\put(35,10){\vector(0,1){60}}
\put(38,78){\large $[\varphi]$}
\put(5,40){\vector(1,0){110}}
\put(113, 33){\large $\varphi$}
\put(10,15){\line(1,1){50}}
\put(60,15){\line(1,1){50}}
\put(10,15){\line(1,1){50}}
\multiput(60,15)(0,5){10}{\line(0,1){4}}
\put(63,33){\large $\pi$}
\put(38,65){\large $\pi$}
\put(38,33){\large 0}
\end{picture}
\vspace{5cm}
\begin{center}
Fig.4
\end{center}
\newpage
\begin{picture}(120,80)
\put(7.5,40){\vector(1,0){110}}
\put(117,35){\large $\varphi$}
\put(60,10){\vector(0,1){60}}
\put(63,70){\large $4\pi j_Smd/3en$}
\put(63,35){\large 0}
\put(48,65){\large $C_1\pi$}
\put(48,50){\large $C_2\pi$}
\put(46,30){\large $-C_2\pi$}
\put(46,15){\large $-C_1\pi$}
\put(60,30){\line(3,2){52.5}}
\put(7.5,15){\line(3,2){52.5}}
\put(7.5,37){\large $-\pi$}
\put(112.5,37){\large $\pi$}
\end{picture}
\vspace{5cm}
\begin{center}
Fig.5a
\end{center}
\newpage
\begin{picture}(120,80)
\put(10,20){\vector(1,0){100}}
\put(60,5){\vector(0,1){60}}
\put(60,50){\line(2,-1){24}}
\put(84,38){\line(2,1){16}}
\put(60,50){\line(-2,-1){24}}
\put(36,38){\line(-2,1){16}}
\put(61,50){$2(A_1-A_2)\pi-A_1\varphi$}
\put(90,36){$A_1\varphi$}
\put(61,17){0}
\put(100,17){$\pi$}
\put(110,15){$\varphi=2\pi\Phi/\Phi_0$}
\put(20,17){$-\pi$}
\put(61,60){$I_{max}$}
\end{picture}
\vspace{5cm}
\begin{center}
Fig.6
\end{center}


\begin{references}
\bibitem{FN1} At temperatures close to the superconducting phase transition
temperature $T_c$ this assumption is justified under
quite general conditions. Furthermore, in the case of
a factorizable pairing potential $ V(\bbox{p},\bbox{p}')=
V_0\psi(\bbox{p})\psi(\bbox{p}')$ the above assumption is correct at
any temperature below $T_c$.  We also assume that singlet pairing is
described by the one-component order parameter. Our analysis can be easily
extended  to the two-component order parameter and accidental degeneracy cases
(the latter corresponds to approximately the same value $T_c$ for two types of
pairing). Then one has
$\Delta(\hat{\bbox{p}})=\eta_1(x)\psi_1(\hat{\bbox{p}})+
\eta_2(x)\psi_2(\hat{\bbox{p}})$.
\bibitem{FN2} The dependence $D(\bbox{p}) \propto p_x^2/p_F^2$ (i.e. $k=2$) is
usually the most relevant one for tunnel junctions with $D \ll 1$.
\bibitem{FN3} One can easily check that the subgap current does not vanish also
if the nodal line of the order parameter of one superconductor comes through
intersection points of the nodal lines for the other.
\end{references}
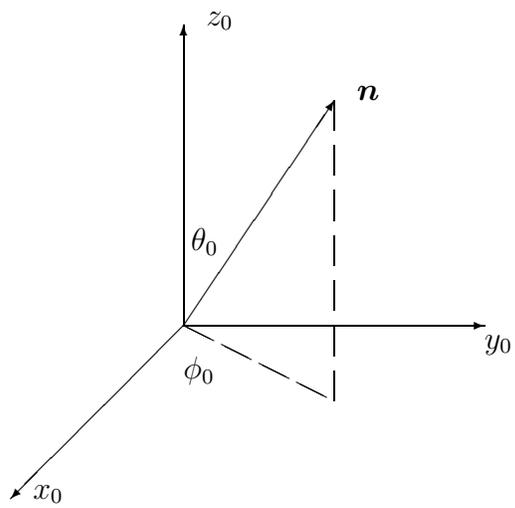
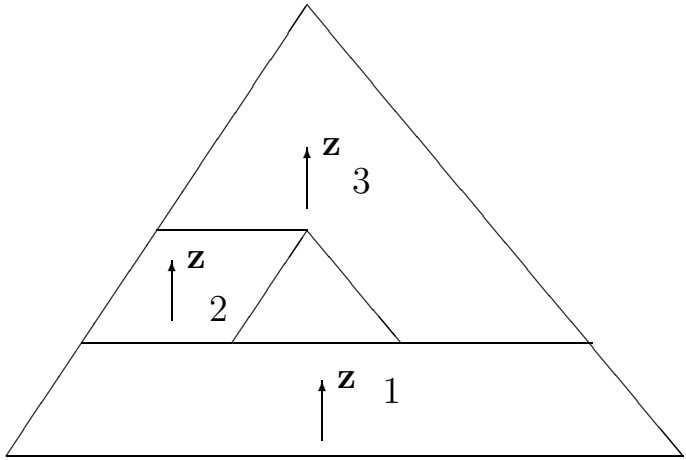
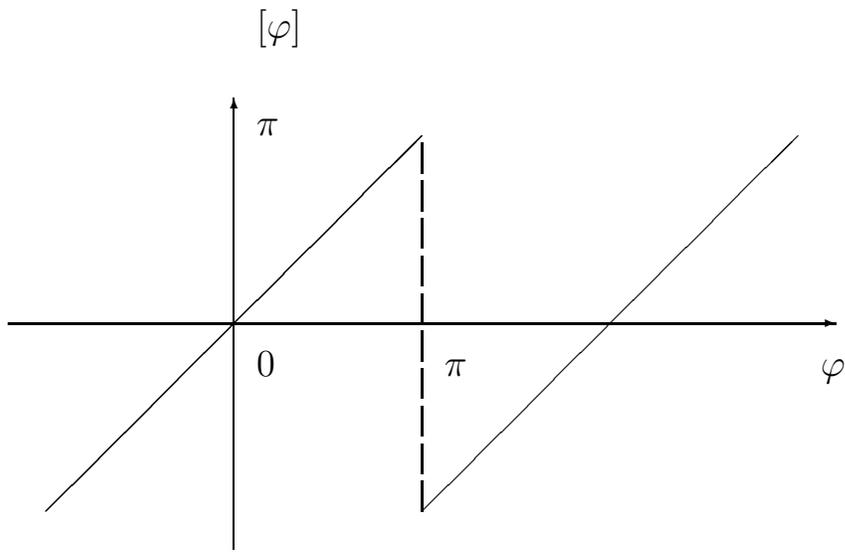
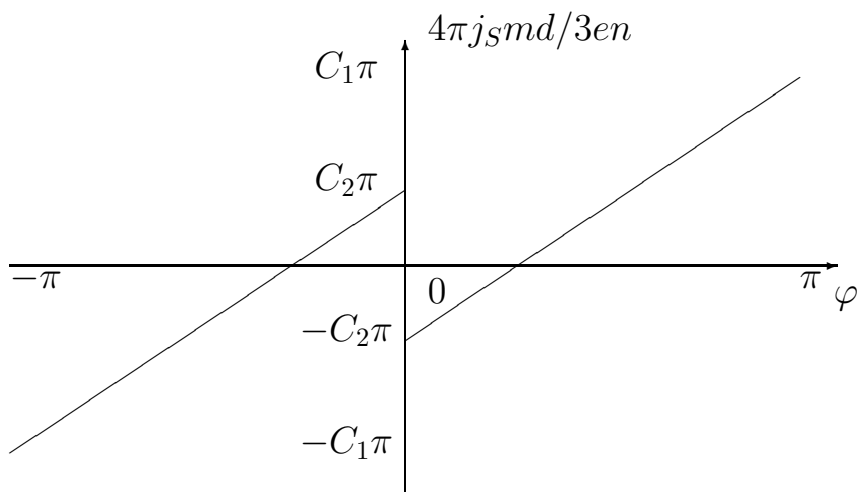
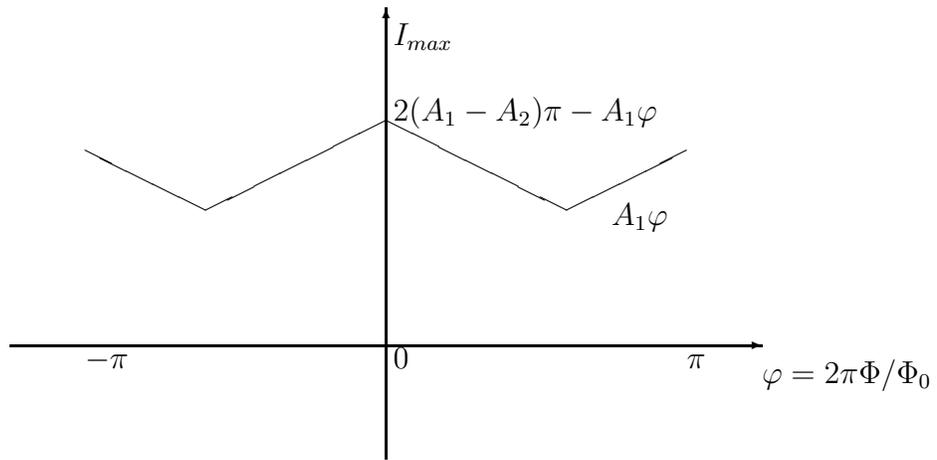
\end{document}